%% file: main.tex
\documentclass[reprint, superscriptaddress,nofootinbib, aps]{revtex4-2}
\input{preamble}

\usepackage{soul} %

\begin{document}
    \title{Distance learning from projective measurements as an information-geometric probe of many-body physics}

	\author{Oleksii Malyshev}	
	\email{omalyshe@ethz.ch}
        \affiliation{Institute for Theoretical Physics, ETH Z\"urich, CH-8093 Z\"urich, Switzerland}

        \author {Simon M. Linsel}
        \affiliation{Department of Physics and Arnold Sommerfeld Center for Theoretical Physics (ASC),
Ludwig-Maximilians-Universit\"at M\"unchen, Theresienstr. 37, M\"unchen D-80333, Germany}
        \affiliation{Munich Center for Quantum Science and Technology (MCQST), Schellingstr. 4, D-80799 M\"unchen, Germany}

        \author {Fabian Grusdt}        
        \affiliation{Department of Physics and Arnold Sommerfeld Center for Theoretical Physics (ASC),
Ludwig-Maximilians-Universit\"at M\"unchen, Theresienstr. 37, M\"unchen D-80333, Germany}
        \affiliation{Munich Center for Quantum Science and Technology (MCQST), Schellingstr. 4, D-80799 M\"unchen, Germany}
        
        \author {Annabelle Bohrdt}            
        \affiliation{Department of Physics and Arnold Sommerfeld Center for Theoretical Physics (ASC),
Ludwig-Maximilians-Universit\"at M\"unchen, Theresienstr. 37, M\"unchen D-80333, Germany}
\affiliation{Munich Center for Quantum Science and Technology (MCQST), Schellingstr. 4, D-80799 M\"unchen, Germany} 

        \author {Eugene Demler}
        \affiliation{Institute for Theoretical Physics, ETH Z\"urich, CH-8093 Z\"urich, Switzerland}
        
        \author{Ivan Morera}
        \affiliation{Institute for Theoretical Physics, ETH Z\"urich, CH-8093 Z\"urich, Switzerland}	

\begin{abstract} 
The ability of modern quantum simulators---both digital and analogue---to generate large ensembles of single-shot projective ``snapshots'' has opened a data-rich avenue for the study of quantum many-body systems.
Unsupervised machine learning analysis of such snapshots has gained traction, with numerous works reconstructing phase diagrams by learning and clustering low-dimensional representations of quantum states.
Here, we forgo such representation learning in favour of \emph{distance} learning: we infer the pairwise distances between quantum states---already sufficient for clustering---directly from snapshots.
Specifically, we use a single neural discriminator to estimate Csiszár $f$-divergences---statistical distances between distributions---in an unsupervised manner.
The resulting clusters reveal regimes with different dominant correlations, often coinciding with, but not limited to, conventionally defined phases of matter. 
Beyond phase-diagram exploration, we connect the infinitesimal limit of the inferred divergences to the Fisher information metric and analyse its finite-size scaling.
This yields critical exponents of the discovered transitions and enables snapshot-based analysis of universality classes.
We apply distance learning to a diverse set of systems characterised by conventional local order parameters (1D transverse-field and 2D classical Ising models), non-local topological order (extended toric code), and higher-order correlations (fermionic $t$--$J$ model on a triangular lattice). 
In all cases, we correctly recover boundaries between distinct correlation regimes and, where applicable, quantitatively match established critical behaviour. 
Finally, we show that distances to suitably chosen reference snapshot distributions help identify the dominant correlations within the discovered clusters, positioning distance learning as a versatile information-geometric probe of quantum many-body physics.
\end{abstract}

\maketitle

\section{Introduction}\label{sec:intro}
\begin{figure*}
	\centering
	\includegraphics[width=\linewidth]{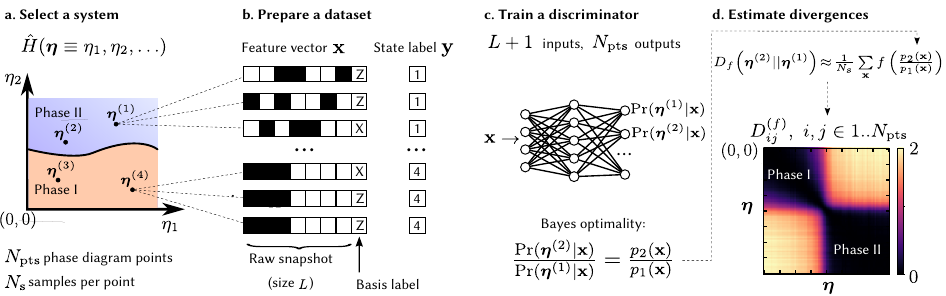}
	\caption{The pipeline for discriminative distance learning. \textbf{a.} We sample the phase diagram by performing projective measurements on the ground state of a studied system (potentially in several bases). \textbf{b.} We prepare the classification dataset, where input vectors are the measurement results (potentially appended with the basis label), and the class labels indicate which phase diagram point the snapshot comes from. \textbf{c.} We train a classifier neural network to predict which phase diagram point a snapshot comes from. \textbf{d.} We use the \emph{posterior} probability ratios produced by the trained classifier to estimate pairwise $f$-divergences between phase diagram points; to that end, we use an importance sampling scheme as described in Section~\ref{sec:discriminative_f_divergence_estimation}.}
	\label{fig:method_pipeline}
\end{figure*}
An important experimental milestone of the past two decades is the development of analogue and digital programmable quantum simulators, accessing quantum systems of ever-increasing size with an ever-growing toolbox of available probes~\cite{georgescu_quantum_simulation,altman_quantum_simulators}.
Remarkably, modern quantum-simulation platforms---including ultracold atoms, Rydberg arrays, superconducting circuits, and trapped ions---are capable of performing projective measurements in their native basis, thus producing snapshots from the prepared quantum state according to the Born rule~\cite{georgescu_quantum_simulation,altman_quantum_simulators, bloch_optical_lattices,schafer_quantum_simulation_tools}.
This provides substantially more information than traditional measurements of local observables and correlation functions, yet also poses the challenge of isolating the physically meaningful structure---especially when studying strongly correlated quantum many-body systems.
Thus, there is a timely need for new tools to harness the data-rich nature of snapshots and guide exploration of the underlying physics.

In this regard, machine learning approaches have garnered particular attention due to their ability to robustly and automatically extract relevant features from data.
Numerous works used unsupervised machine learning to explore phase diagrams directly from raw snapshots, recovering boundaries of theoretically known phases across systems with conventional and topological order~\cite{wang_ising_pca,hu_critical_examination, rodriguez_diffusion_maps,lidiak_diffusion_maps, sadoune_pollet_svm, coles_rydberg_pca}.
In a typical setup, snapshots are produced at different points of the quantum many-body phase diagram.
One compresses each ensemble of snapshots into a low-dimensional representation of the corresponding state and clusters these representations.
Distinct clusters are then interpreted as regimes characterised by qualitatively different physics.
The representation learning techniques considered so far include principal component analysis~\cite{wang_ising_pca,hu_critical_examination}, diffusion maps~\cite{rodriguez_diffusion_maps,lidiak_diffusion_maps}, as well as custom-designed algorithms~\cite{sadoune_pollet_svm,coles_rydberg_pca}. 
However, despite the reported successes of such methods across various systems, the problem of which representation---if any---best suits a given system remains without a systematic understanding.

Here, we bypass representation learning altogether and resort to \emph{distance} learning.
First, we note that clustering algorithms can operate directly on the matrix of pairwise distances between data points, without requiring an explicit low-dimensional representation.
Second, we argue that natural distances between snapshot ensembles are statistical distances between the underlying Born rule probability distributions.
We therefore use a single self-supervised discriminator (classifier) neural network to estimate Csiszár $f$-divergences---a well-studied family of statistical distances rooted in information geometry~\cite{polyanskiy_information_theory_textbook}---between all pairs of the measured states directly from snapshots (see Fig.~\ref{fig:method_pipeline}).
This is in contrast to the initial works on distance learning which used Siamese networks~\cite{patel_siamese}, contrastive learning~\cite{han_simclr}, and pairwise support vector machine intercepts~\cite{sadoune_pollet_svm} to produce model-specific, data-driven distance surrogates. 

Importantly, distance learning is not restricted to ground states: the measured states may be excited, thermal, non-equilibrium, or---more generally---any set of experimentally prepared quantum states.
We further show that distance learning naturally incorporates measurements performed in different bases, is robust to the choice of training hyperparameters, and requires modest computational and sample budgets.

Beyond clustering, we consider the \emph{susceptibility} of the learnt $f$-divergences, known to yield the Fisher information metric on the phase-diagram manifold~\cite{polyanskiy_information_theory_textbook}.
We connect this metric to the standard probes of criticality: heat capacity for classical Gibbs states, and fidelity susceptibility for quantum ground states~\cite{gu_fidelity_approach_review,albuquerque_fidelity_susceptibility}.
Importantly, both these quantities diverge at phase transitions, with critical exponents tightly linked to that of the correlation length~\cite{gu_fidelity_approach_review,albuquerque_fidelity_susceptibility, arnold_connection_to_fisher}.
We extract these exponents by analysing the finite-size scaling of the inferred Fisher metric.
This makes it possible to study universality classes---and hence emergent universal behaviour across microscopically distinct systems---directly from experimental snapshot data and on a purely information geometric footing.

To demonstrate the broad utility of distance learning, we apply it to a range of systems recently discussed within the context of quantum simulation (see Fig.~\ref{fig:studied_systems}).
First, we study systems characterised by conventional local order parameters: the paradigmatic transverse-field Ising model in one dimension and the classical two-dimensional Ising model.
In both cases we identify the transition point and extract the critical behaviour via finite-size scaling, recovering the established critical exponents.
Second, to probe regimes with non-local topological order, we map the phase diagram of the extended toric code model on different lattice geometries, where no universally acknowledged order parameter is known~\cite{linsel_percolation,linsel_independent_confinement}.
Interestingly, we resolve a distinct region adjacent to the first-order line at high external fields, suggestive of a supercritical-like regime in a liquid--gas phase diagram, and previously only touched upon in this model~\cite{xu_one_form_symmetries}.  
Third, we address physics governed by higher-order correlations.
Specifically, we consider snapshots generated from the doped fermionic $t$--$J$ model on a triangular lattice hosting either the recently studied Nagaoka polaron phase or kinetically induced bound states~\cite{morera_nagaoka,he2023itinerantmagnetismtriangularlattice,Chen2025,morera_t_j_ladders,qiao_browayes_nagaoka}.
In the Nagaoka polaron setting, distance learning produces clusters tracking the position of the peak in the static spin structure factor.
For the kinetically induced bound states, we show that the inferred cluster boundaries quantitatively match the threshold exchange strengths at which different bound states form.

By construction, however, distance learning is mainly a tool of exploration rather than interpretation.
It discovers which phase diagram regions differ qualitatively, while explaining physics of each distinct region requires further analysis and domain knowledge.
Nonetheless, we show that it can provide system-specific insights into the nature of the discovered clusters too.
For the Nagaoka regime of the $t$--$J$ model on the triangular lattice, we estimate distances to synthetic reference data to show that the relevant structure is encoded in higher-order correlations between dopants and spins.
For the composite bound states in the same model, we show that representations learnt by the discriminator generalise across particle-number sectors, enabling comparison between non-overlapping quantum states. 
We thus believe that distance learning can serve as a natural and practical tool for navigating the wealth of information contained in the quantum simulator snapshots.

\section{Methods}\label{sec:method} 
\subsection{High-level overview}\label{sec:high_level_overview}
Fig.~\ref{fig:method_pipeline} presents a high-level overview of distance learning.
For concreteness, we assume access to \Npoints{} ground states of a Hamiltonian \Hhatparam{} corresponding to different values of its control parameters \etavec{} (e.g. external fields, on-site repulsion, doping, etc.).
However, the same analysis applies to any set of quantum states, either parameterised (e.g., Gibbs ensembles, time-evolved states, etc.) or not.
At each point $\etaveci$ ($i \in 1..\Npoints$) we acquire \Nsnapshots{} single-shot projective measurements (``snapshots'') represented as vectors $\xvec$ (e.g., measurements of particle number, spin projection, etc.).
Each \xvec{} is drawn from the distribution $p_i(\xvec) \propto \left|\psi_i(\xvec)\right|^2$ related to the wave function via the Born rule---we term it \emph{Born distribution}.
In total, we collect $\Nsnapshots \cdot \Npoints$ snapshots (Fig.~\ref{fig:method_pipeline}a).
Our goal is to estimate pairwise statistical distances $D(p_i || p_j)$ between these snapshot distributions and assemble them into a matrix $D_{ij}$ ($i, j \in 1..\Npoints$).

Each ensemble of snapshots is a finite sample from a high-dimensional distribution, and \emph{per se} provides only a crude empirical proxy for $p_i(\xvec)$ for a given snapshot $\xvec$. 
In addition, ensembles corresponding to different quantum states often share only a small common support.
As a result, na\"ively estimating distances as the ``overlap'' between pairs of snapshot ensembles yields unsatisfactory results (see Appendix~\ref{sec:naive_overlaps} for more details).

We therefore leverage a discriminator neural network to obtain more accurate distance estimates---we provide theoretical justification in Section~\ref{sec:discriminative_f_divergence_estimation}.
Given a snapshot $\xvec$, the discriminator  produces \Npoints{} \emph{posterior} probabilities $\Pr\!\left( i|\xvec\right)$, i.e., the probability that $\xvec$ was sampled from the distribution $p_i$.
Training a discriminator is equivalent to solving a \emph{classification} problem: we treat each snapshot distribution  (each \etaveci{}) as a separate class, and ask the network to predict the label $i$ given $\xvec$.

Accordingly, we prepare a self-supervised classification dataset: each snapshot $\xvec$ is paired with a class label $i$ indicating the phase diagram point it comes from (Fig.~\ref{fig:method_pipeline}b).
If snapshots come from multiple bases, we append a basis label $\alpha$ to each snapshot and form an augmented input $(\xvec, \alpha)$.
We then train a neural network classifier on this dataset (Fig.~\ref{fig:method_pipeline}c), a standard task in modern machine learning (see Appendix~\ref{sec:nn_architecture_and_training} for details on the classifier architecture and training procedure).

Finally, we use these posterior probabilities in a Monte Carlo estimator for $D(p_i  \Vert  p_j)$ (see below). 
Estimating all pairwise distances yields the matrix $D_{ij}$ (Fig.~\ref{fig:method_pipeline}d), which we then use for phase clustering and the analysis of critical behaviour.

\subsection{Discriminative $f$-divergence estimation}\label{sec:discriminative_f_divergence_estimation}
A specific family of distances we reconstruct are \emph{Csiszár $f$-divergences}.
An $f$-divergence between two probability distributions $p$ and $q$ reads as:
\begin{equation}\label{equ:f_div_def}
    \fdivarg{q}{p} \coloneqq \mathexp_{\xvec \sim p} 
    \left[f\!\left(\frac{q(\xvec)}{p(\xvec)}\right)\right].
\end{equation}
Here $\mathexp_{\xvec \sim p} $ denotes the mathematical expectation taken over the distribution $p(\xvec)$, while $f(\cdot)$ is a convex function with $f(1)=0$, subject to a few mild regularity constraints~\cite{polyanskiy_information_theory_textbook}.
Csiszár $f$-divergences provide a unified perspective on various statistical divergences known in the literature.
For example, the well-known total variation (TV), Kullback-Leibler (KL), and squared Hellinger divergences correspond to $f(t) = \frac{1}{2}\left|t - 1\right|$, $f(t) = t\ln t$, and $f(t) = \left(\sqrt{t} - 1\right)^2$, respectively~\cite{polyanskiy_information_theory_textbook}.

We use $f$-divergences for two reasons.
First, they form a natural, well-studied family of divergences with a direct link to the Fisher information, and thus to the underlying information geometry (see below).
Second, Eq.~\eqref{equ:f_div_def} can be used to construct a \emph{practical} importance sampling Monte Carlo estimator for $\fdivarg{q}{p}$:
\begin{equation}\label{equ:crude_importance_sampling_estimator}
    \fdivarg{q}{p} = \frac{1}{2 \Nsnapshots} \sum_{\xvec \sim  p + q} w(\xvec) \cdot f\left(r(\xvec) \right).
\end{equation}
Here the summation is performed over the samples in the training dataset which come either from $p$, or from $q$; $r(\xvec) = q(\xvec) / p(\xvec)$ is the so-called \emph{density ratio}; and $w(\xvec) = \frac{2}{1 + r(\xvec)}$ is the importance weight. 
We prove the validity of this estimator in Appendix~\ref{sec:importance_sampling_derivation}.

Crucially, this estimator requires knowledge only of the density ratio $r(\xvec)$.
In principle, $r(\xvec)$ can be obtained by separately estimating the normalised probabilities $p(\xvec)$ and $q(\xvec)$.
We, however, estimate $r(\xvec)$ \emph{directly}, which is known to be an easier statistical task~\cite{density_ratios_in_ml}.
To that end, we use the posterior probabilities $\Pr\!\left(p|\xvec\right)$ and $\Pr\!\left(q|\xvec\right)$ produced by a trained discriminator neural network.
Assuming that equal numbers of snapshots from $p$ and $q$ enter the dataset, and the network was trained to reach \emph{Bayesian optimality}, the ratio of posterior probabilities $\frac{\Pr\!\left(q|\xvec\right)}{\Pr\!\left(p|\xvec\right)}$ is known to equal $r(\xvec)$~\cite{devroye_bayes_optimal_textbook}.

In Appendix~\ref{sec:importance_sampling_derivation} we extend the estimator~\eqref{equ:crude_importance_sampling_estimator} to (i) non-uniform number of snapshots produced per quantum state and (ii) measurements performed in multiple bases.
We model each basis $\alpha$ by a conditional snapshot distribution $p(\xvec|\alpha)$, and denote by $\pi_\alpha^{(p)}$ the fraction of snapshots measured in the basis $\alpha$ for the state $p$ ( $\sum_\alpha \pi_\alpha^{(p)} = 1$).
The resulting total $f$-divergence is a weighted sum of per-basis $f$-divergences:
\begin{equation}\label{equ:multiple_bases_divergence_equivalence}
    \fdivarg{q}{p} = \sum_{\alpha} \pi_\alpha^{(p)} \cdot \fdivarg{q(\xvec|\alpha)}{p(\xvec|\alpha)}.
\end{equation}
\subsection{Phase diagram clustering}
While the described framework is applicable to the estimation of \emph{any} $f$-divergence, in the distance-learning applications considered below we use the squared Hellinger divergence $H^2 (p, q)\coloneqq \sum_{\xvec}(\sqrt{q(\xvec)} - \sqrt{p(\xvec)})^2$.
We see it as a natural choice for Born distributions, since for real non-negative wavefunctions $H^2 (p, q)$ is related to the overlap between the states as $\braket{p|q} = 1 - \frac{H^2 (p, q)}{2}$.
Its square root also satisfies the axioms of a true distance metric, and thus naturally lends itself to distance-based clustering. 
There exist, however, other $f$-divergences whose square roots are true distance metrics too---e.g. Jensen-Shannon and Le Cam~\cite{polyanskiy_information_theory_textbook}---and in Appendix~\ref{sec:nn_architecture_and_training} we discuss that predictions yielded by these divergences are qualitatively and quantitatively similar to those obtained with the Hellinger divergence.

To cluster the phase diagram, we take square roots of the estimated divergences to obtain the matrix $\sqrt{D_{ij}}$ of pairwise Hellinger distances $H(p_i, p_j)$.
We then apply the hierarchical density-based spatial clustering algorithm (HDBSCAN)~\cite{campello_hdbscan} to the resulting distance matrix.
Compared to more traditional clustering algorithms such as K-means and Gaussian mixture models, this algorithm is advantageous for three reasons.
First, it can cluster data points based only on the matrix of precomputed distances between them.
Second, it makes no assumptions about either the shape of clusters or their number and can assign a ``no decision'' label to a data point not clearly belonging to any of the found clusters.
Finally, HDBSCAN provides a measure of confidence in its cluster assignment for every data point.
Thus, using HDBSCAN enables quantitative exploration of the phase diagram with minimal prior assumptions.

\subsection{$f$-divergence susceptibility and its critical scaling}\label{sec:f_divergence_susceptibility}
\begin{figure*}
	\centering
	\includegraphics[width=\linewidth]{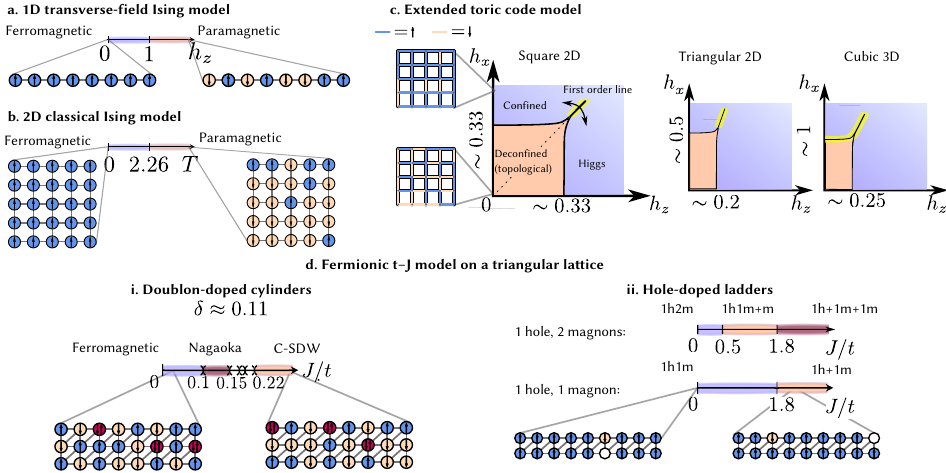}
    \caption{Schematic phase diagrams and representative snapshots for the studied systems. \textbf{a.} Transverse-field Ising model hosts two phases---ferromagnetic (FM) and paramagnetic (PM)---separated by a quantum critical point at $h_z/J = 1$. \textbf{b.} Similarly, the classical ferromagnetic Ising model hosts a ferro- and paramagnetic phase separated by the critical point at $T \approx 2.26$. \textbf{c.} The extended toric code hosts a topologically ordered deconfined phase at low external fields, which transitions into a confined (Higgs) phase as $h_x$ ($h_z$) grows. The last two phases can be adiabatically connected by a path around a first-order critical line (highlighted with yellow)~\cite{linsel_percolation}. The phase diagrams for triangular (2D) and cubic (3D) lattices are qualitatively similar and are adapted from Ref.~\cite{linsel_independent_confinement}. \textbf{d.} (i) the fermionic $t$--$J$  model on a triangular cylinder at small doublon doping ($\delta \approx 0.11$) hosts a multitude of regions primarily characterised by the position of the peak in the spin structure factor $S_{zz}(\mathbf{q})$. One can nevertheless distinguish a traditional ferromagnetic phase at low superexchange, a kinetically induced Nagaoka phase at $J/t$ from $0.1$ to $0.15$, and a commensurate spin-density wave at $J > 0.22$~\cite{morera_nagaoka}; (ii) the fully spin-polarised state of the fermionic $t$--$J$  model on a triangular ladder doped with holes and magnons hosts various multiparticle bound states~\cite{morera_t_j_ladders}.}
	\label{fig:studied_systems}
\end{figure*}
While identifying distinct clusters in the phase diagram is an important step of the initial analysis, it is also desirable to gain better understanding of the nature of the identified transitions. 
To that end, we consider the $f$-divergence susceptibility for a family of parametrised probability distributions $p_{\eta}(\xvec)$:
\begin{equation}\label{equ:f_divergence_susceptibility}
    \chi(\eta) = \lim_{\Delta \eta \to 0} \frac{\fdivarg{p_{\eta + \Delta \eta}}{p_{\eta}}}{\left|\Delta \eta\right|^2}.
\end{equation}
It is known to be proportional to the classical Fisher information $I(\eta)$, which is an information-geometric object measuring how distributions vary on a parametrised manifold~\cite{polyanskiy_information_theory_textbook}:
\begin{equation}\label{equ:f_divergence_susceptibility_fisher}
    \chi(\eta)= \frac{1}{2} f^{\prime\prime}(1) I(\eta), \quad I(\eta) = \mathbb{E}_{\xvec \sim p_{\eta}} \left[ \left( \partial_{\eta} \log p_{\eta}(\xvec) \right)^2 \right].
\end{equation}
In the case of multi-parameter families, $I(\etavec)$ and $\chi(\etavec)$ are straightforwardly promoted to matrices~\cite{polyanskiy_information_theory_textbook}.

This connection is appealing since the Fisher information is naturally related to standard susceptibilities studied in statistical physics.
For instance, for thermal states following the Gibbs distribution $p_{T}(\xvec) \propto e^{-H(\xvec)/T}$ with Hamiltonian $H(\xvec)$, the Fisher information along the temperature direction is related to the heat capacity as follows (see Appendix~\ref{sec:fisher_information_thermal_states} for a derivation):
\begin{equation}\label{equ:fisher_information_heat_capacity}
    I(T) = \frac{\mathrm{Var}_{\xvec \sim p_{T}}\!\left[H(\xvec)\right]}{T^4} = \frac{C_V}{T^2}.
\end{equation}
Importantly, studying critical scaling of such susceptibilities is a well-established approach to characterising phase transitions and universality classes.

Assuming $\fdivarg{p_{\eta + \Delta \eta}}{p_{\eta}}$ was calculated using measurements in several bases as per Eq.~\eqref{equ:multiple_bases_divergence_equivalence}, we can further connect $\chi(\eta)$ to the \emph{quantum} Fisher information---also known as fidelity susceptibility $\chi_{\mathrm F}(\eta) \coloneqq -\frac{1}{2}\left.\frac{\partial^2}{\partial(\Delta\eta)^2}\,
\left|\braket{\psi_\eta | \psi_{\eta+\Delta\eta}}\right|^2\right|_{\Delta\eta=0}$ ~\cite{gu_fidelity_approach_review}.
By differentiating Eq.~\eqref{equ:multiple_bases_divergence_equivalence} twice with respect to $\Delta \eta$, it follows that $\chi(\eta)$ is proportional to the weighted sum of per-basis classical Fisher information $I_{\alpha}(\eta)$:
\begin{equation}
    \chi(\eta) \sim \sum_\alpha \pi_\alpha^{(p_{\eta})} I_\alpha(\eta).
\end{equation}
Ref.~\cite{lu_qfi_via_random_cfi} showed that for measurement bases generated by Haar-random local rotations---i.e., sampled uniformly from the space of locally rotated bases---this average is proportional to $\chi_{\mathrm F}$.
Therefore, as the number of available bases grows, $\chi(\eta)$ approaches $\chi_{\mathrm F}$.

Similarly to the heat capacity, $\chi_{\mathrm F}$ serves as a probe of criticality for \emph{quantum} phase transitions.
As the parameter $\eta$ is varied across a continuous phase transition, $\chi_{\mathrm F}$ typically diverges.
This reflects the critical sensitivity of the ground state---and hence its Born distribution---to an infinitesimal parameter change. 
The scaling of this divergence has been extensively studied in the past~\cite{gu_fidelity_approach_review,albuquerque_fidelity_susceptibility}. 
In particular, for a system of dimension $d$ and linear size $L$, the fidelity-susceptibility density $\chi_{\mathrm F}^{(d)} \coloneqq \chi_{\mathrm F}/L^d$ obeys hyperscaling with a critical exponent $\alpha_{\mathrm F}$ fixed by the correlation-length exponent $\nu$~\cite{gu_fidelity_approach_review,albuquerque_fidelity_susceptibility}:
\begin{equation}\label{equ:fidelity_susceptibility_critical_exponent_relation}
    \chi_{\mathrm F}^{(d)} \propto \left|\eta - \eta_{\mathrm c}\right|^{-\alpha_{\mathrm F}} \quad \alpha_{\mathrm F} = 2 - d \nu .
\end{equation}

If $\eta^{(i)}$ are sampled on a regular grid with spacing $\Delta \eta$, the finite-difference approximation of $\chi(\eta^{(i)})$ can be easily extracted from the values contained in the matrix $D_{ij}^{(f)}$.
One can then perform finite-size scaling analysis of $\chi(\eta)$ and extract the correlation-length and fidelity-susceptibility critical exponents of the identified transition (see Section~\ref{sec:tfim_section} and Section~\ref{sec:classical_2d_ising}).

In practice, implementing Haar-random measurements is challenging. 
Nevertheless, if the transition is visible in at least one of the chosen bases, we expect $\chi(\eta)$ to diverge at the transition and provide a lower bound on the critical scaling of the fidelity susceptibility. 
Moreover, if the wavefunction is real and non-negative in a given basis, measurements in this basis suffice to reconstruct $\chi_{\mathrm F}$ and perform finite-size scaling, as exemplified by our transverse-field Ising study (Section~\ref{sec:tfim_section}). 

\section{Results}

\subsection{Transverse-field Ising model}\label{sec:tfim_section}
\subsubsection{Model}
\begin{figure*}
	\centering
	\includegraphics[width=\linewidth]{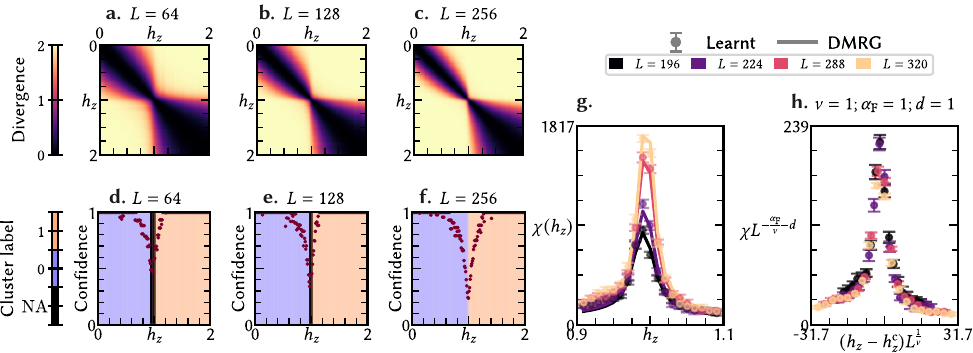}
	\caption{Discriminative distance learning of the transverse-field Ising model phase diagram. \textbf{a--c.} Matrices of pairwise Hellinger divergences between ground states of the TFIM for different system sizes $L$.  \textbf{d--f.} Phase clustering results obtained by applying the HDBSCAN algorithm to the divergence matrices in panels a--c. Shaded areas of different colours indicate different clusters; black shaded areas indicate phase diagram points not attributed to any cluster; red points show the HDBSCAN clustering confidence for every phase diagram point. \textbf{g.} Finite-difference approximation to the Hellinger divergence susceptibility $\chi(h_z)$ defined in Eq.~\eqref{equ:f_divergence_susceptibility} for different system sizes $L$. Error bars are obtained via repeating distance learning for 19 random initialisations of the discriminator. \textbf{h.} Finite-size scaling collapse of the Hellinger divergence susceptibility for different system sizes $L$.}
	\label{fig:tfim_results}
\end{figure*}
As a benchmark application of distance learning, we consider the paradigmatic one-dimensional (1D) transverse-field Ising model (TFIM).
This model describes a chain of $L$ interacting spins subjected to an external transverse magnetic field, and is governed by the Hamiltonian
\begin{equation}\label{equ:tfim_hamiltonian}
\Hhat_{\rm TFIM} = -J\sum_{i}\hat{\sigma}_i^x \hat{\sigma}_{i+1}^x - h_z \sum_{i} \hat{\sigma}_i^z,
\end{equation}
with $J$ and $h_z$ denoting the spin-spin interaction and magnetic field strengths, respectively.

For $J>0$, the spin-spin interaction favours a symmetry-broken ferromagnetic state aligned along $+x$ or $-x$ directions.
By contrast, the external magnetic field polarises the spins along the $+z$ direction, resulting in a disordered paramagnetic state along $x$.
Depending on the ratio between $J$ and $h_z$, one of two competing orders wins.
The transition from the ferro- to paramagnetic phase is known to occur at $h_z/J = 1$~\cite{sachdev_qpts}.

\subsubsection{Phase diagram mapping}
We apply distance learning to chains of sizes $L = \{64, 128, 256\}$.
For each system size we consider 201 values of $\eta \coloneqq h_z/J$ taken uniformly on the $[0, 2]$ interval and use the density matrix renormalisation group (DMRG) algorithm to obtain the corresponding ground states.
We sample 2000 snapshots in the $x$ basis for each ground state and calculate the $f$-divergence between each pair of states for a given $L$.
We present the resulting matrices of pairwise divergences in Fig.~\ref{fig:tfim_results}a--c.

These matrices exhibit a block structure, with two block-diagonal regions of low divergence and off-diagonal regions of high divergence.
We associate the observed dark regions with the two known phases of the TFIM.
As $L$ grows, the ``waist'' of the crossover region between two phases becomes thinner.
We relate this region to the critical region around the phase transition point, where the correlation length exceeds the system size, and which is expected to shrink with increasing $L$.
Thus, the raw output of the discriminative $f$-divergence estimation already contains clear evidence of a phase transition and the existence of distinct phases in the system.
Let us remark that a much simpler approach---estimating overlaps between states directly from snapshots---yields zero for almost all pairs of states, as discussed in Section~\ref{sec:high_level_overview} (see Appendix~\ref{sec:naive_overlaps} for more details). 
 
Further HDBSCAN clustering consistently splits the phase diagram in two clusters for all considered system sizes as shown in Fig.~\ref{fig:tfim_results}d--f, where we highlight the regions belonging to different clusters with different colours.
The points close to the thermodynamic phase transition are not assigned to any cluster, which we attribute to them falling into the critical region caused by the finite-size effects (black shaded areas in Fig.~\ref{fig:tfim_results}d--f). 
As $L$ increases this unclassified region shrinks, reflecting the expected sharpening of the crossover between the phases.  
Additionally, the HDBSCAN confidence decreases for points close to the critical region.

\subsubsection{Critical scaling of the $f$-divergence susceptibility}
Having identified the two phases of the TFIM, we now examine whether our approach can accurately capture the critical exponents of the transition.
In the $x$ basis the TFIM Hamiltonian is stoquastic, and thus the ground states are real and non-negative.
As a result, measurements in the $x$ basis alone suffice to reconstruct the critical exponent $\alpha_{\mathrm F}$, since the quantum Fisher information coincides with the classical one.
In Fig.~\ref{fig:tfim_results}g we show the finite-difference approximation to $ \chi(\eta)$ calculated for several system sizes.
As expected, it peaks around the phase transition point, and, as $L$ grows, the peak becomes sharper and higher, while its position shifts closer to the true critical point.
We also plot the exact values of $\chi(\eta)$ obtained by estimating the Hellinger distances directly from the overlaps between the ground states MPSes.
We find the exact and learnt values to be in close agreement.

Fig.~\ref{fig:tfim_results}h shows density of $\chi(\eta)$ rescaled according to the finite-size scaling ansatz described in Appendix~\ref{sec:fss_ansatz}.
We use the known critical point $\eta_{\mathrm c} = 1$, correlation length exponent $\nu = 1$, and the fidelity susceptibility exponent $\alpha_{\mathrm F} = 1$ for the rescaling~\cite{albuquerque_fidelity_susceptibility,gu_fidelity_approach_review}, and find a good collapse of the curves for different $L$.
As a quantitative alternative to by-eye assessment, we performed an automated finite-size scaling collapse of $\chi(\eta)/L^d$ (see Appendix~\ref{sec:fss_ansatz} for more details).
It yielded the correlation-length exponent $\nu = 1.13 \pm 0.1$ and the susceptibility exponent $\alpha_{\mathrm F} = 0.8 \pm 0.05$.
Both are in good agreement with the theoretically known values and the scaling relation~\eqref{equ:fidelity_susceptibility_critical_exponent_relation}.
Thus, distance learning combines phase-diagram exploration with critical-exponent extraction in a unified pipeline operating directly on projective measurements. 
The underlying learnt observable is a susceptibility with established physical meaning and critical scaling. 
This provides an alternative to earlier machine-learning approaches, which extracted the correlation-length exponent $\nu$ from raw snapshots using data-driven observables such as intrinsic dimension, for which the scaling theory is less well established~\cite{mendes_santos_intrinsic_dimension_scaling}.

\subsection{Ferromagnetic 2D Ising model}\label{sec:classical_2d_ising}
\begin{figure}
	\centering
	\includegraphics[width=\linewidth]{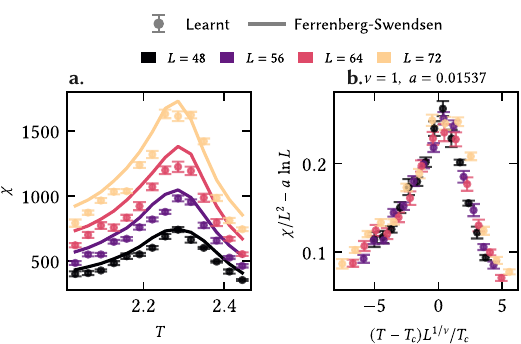}
    \caption{Learnt $f$-divergence susceptibilities for the classical 2D Ising model. \textbf{a.} Finite-difference approximation to the Hellinger divergence susceptibility $\chi(T)$ defined in Eq.~\eqref{equ:f_divergence_susceptibility} for different lattice sizes $L$. Solid lines are produced using Ferrenberg-Swendsen histogram reweighting (see Appendix~\ref{sec:theoretical_heat_capacity_2d_ising}).  Error bars are obtained via repeating distance learning for 8 random initialisations of the discriminator. \textbf{b.} Finite-size scaling collapse of the curves from panel \textbf{a} using the known critical exponents of the 2D Ising universality class. The near-zero value of $a$ indicates that the logarithmic divergence is weakly visible for the system sizes considered and may be further obscured by the finite temperature step $\Delta T$.}
	\label{fig:classic_2d_ising_results}
\end{figure}
We further demonstrate the ability of distance learning to probe critical behaviour by studying the \emph{classical} ferromagnetic 2D Ising model.
This model describes spins residing on the vertices of a 2D square lattice, interacting with their nearest neighbours according to the Hamiltonian
\begin{equation}\label{equ:classical_2d_ising_hamiltonian}
    H_{\mathrm{CFIM}}(\svec) = -J \sum_{\langle i, j \rangle} s_i s_j.
\end{equation}
Here $s_i = \pm 1$ are classical spin variables, and the sum runs over all nearest-neighbour pairs.
The system undergoes a phase transition from a low-temperature ferromagnetic phase to a high-temperature paramagnetic phase at the critical temperature $T_c = \frac{2J}{\ln(1 + \sqrt{2})} \approx 2.269 J$~\cite{baxter_2d_ising_exact_solution}.
Despite its distinct microscopic origin from the 1D TFIM (classical finite-$T$ versus quantum $T=0$), it belongs to the same universality class and shares the same critical exponents~\cite{sachdev_qpts}.
This makes it a natural testbed for the ability of distance learning to capture universal critical behaviour.

We use Monte Carlo sampling with the Wolff update rule~\cite{wolff_update} to generate snapshots at temperatures around the phase transition on square $L\times L$ lattices with $L\in\{48,56,64,72\}$. 
In Fig.~\ref{fig:classic_2d_ising_results}a we show the finite-difference approximation to $\chi(T)$ inferred from the learnt distance matrices.
With solid lines we plot another approximation to $\chi(T)$, obtained with Ferrenberg-Swendsen histogram reweighting~\cite{ferrenberg_swendsen_histogram_reweighting_1,ferrenberg_swendsen_histogram_reweighting_2}, and find both methods in good agreement (see Appendix~\ref{sec:theoretical_heat_capacity_2d_ising} for more details).
Ferrenberg-Swendsen reweighting, however, explicitly requires samples from the Gibbs distribution with a known Hamiltonian, while distance learning is Hamiltonian-agnostic and is applicable to any family of parameterised states.

As discussed in Section~\ref{sec:f_divergence_susceptibility}, for Gibbs ensembles $\chi(T) / L^d$ diverges with the critical exponent of the specific heat. 
In the 2D Ising universality class this exponent is known to be $\alpha=0$~\cite{baxter_2d_ising_exact_solution}.
This means that in fact $\chi(T) / L^d$ diverges logarithmically: $\chi(T) / L^d \propto a\ln L + g((T - T_c) L^{1/\nu})$ for some constant $a$ and scaling function $g(\cdot)$.
In Fig.~\ref{fig:classic_2d_ising_results}b we show the rescaled curves $\chi(T) / L^d$ with the subtraction of the fitted logarithmic term and observe good collapse.
This suggests that distance learning captures critical behaviour and enables analysis of universality across microscopically distinct systems.

\subsection{Extended toric code}
\begin{figure*}
	\centering
	\includegraphics[width=\linewidth]{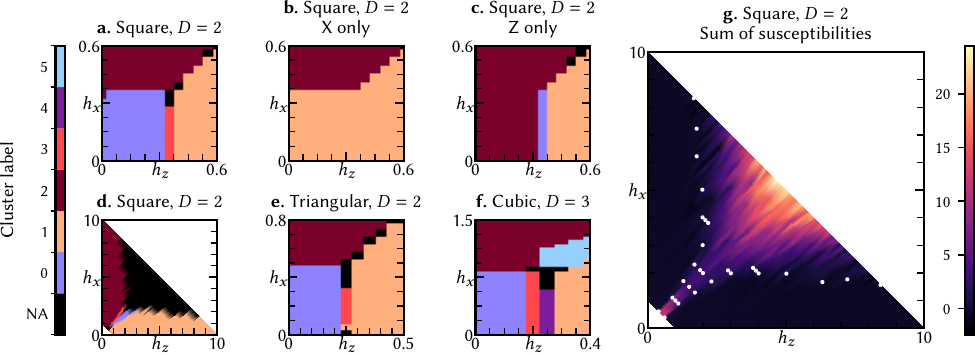}
	\caption{Discriminative distance learning of the extended toric code phase diagram. All phase diagrams are reconstructed from snapshots produced in both $x$ and $z$ bases, unless specified otherwise (figures \textbf{b, c}). \textbf{a.} Phase diagram at small external fields. \textbf{b/c.} Phase diagram at small external fields reconstructed from snapshots produced in $x$/$z$ basis only. \textbf{d.} Phase diagram at large external fields featuring a conjectured supercritical region not attributed to any cluster. \textbf{e.} Phase diagram reconstructed for the toric code on a 2D triangular lattice. \textbf{f.} Phase diagram reconstructed for the toric code on a 3D cubic lattice; each plaquette comprises four spins on a unit-cell face. \textbf{g.} The sum of $\sigma_x$ and $\sigma_z$ susceptibilities at high external fields; the borders of the found clusters from panel \textbf{d} are indicated with white dotted lines.}
	\label{fig:toric_code_results}
\end{figure*}
\subsubsection{Model}
While Ising models provide a controllable benchmark, they have limited relevance for assessing the broader applicability of distance learning.
Their phases are described by a local order parameter, the phase transition is of the conventional symmetry-breaking type, and presence of two phases is already evident after a simple linear transformation applied to the snapshots~\cite{wang_ising_pca}.
Hence, we apply distance learning to a more challenging extended toric code model, which hosts non-local topological order and requires tailored snapshot-based order parameters, introduced only very recently~\cite{linsel_percolation,linsel_independent_confinement,dunnweber_percolation_rg}.

This model describes spins residing on the links of a square 2D lattice of linear size $L$ with periodic boundary conditions ($2L^2$ spins in total); extensions to other lattice geometries and dimensions are considered below. 
The Hamiltonian reads as follows:
\begin{equation}\label{equ:toric_code_hamiltonian}
    \begin{gathered}        
\hat{H}_{\mathrm{TC}}= -\sum_{+} \prod_{l\in +} \hat{\sigma}^{x}_{l}
   - \sum_{\square} \prod_{l\in \square} \hat{\sigma}^{z}_{l} \\
   - h_x \sum_{l} \hat{\sigma}^{x}_{l}
   - h_z \sum_{l} \hat{\sigma}^{z}_{l}.
    \end{gathered}
\end{equation}
The first ``star'' term sums interactions between every quartet of spins sharing the same vertex.
The second ``plaquette'' term sums interactions between every quartet of spins sharing the same plaquette.
Together, these two terms constitute the bare toric code Hamiltonian.
This Hamiltonian describes a paradigmatic system hosting bulk topological order (gapped $\mathbb{Z}_2$ quantum spin liquid), characterised by long-range entanglement and ground state degeneracy~\cite{simon_topological_quantum}.

The third and fourth terms in Eq.~\eqref{equ:toric_code_hamiltonian} couple spins to external magnetic fields along $x$ and $z$ directions respectively.
As the fields increase, the topological order gets disrupted, and the system transitions to trivial phases (see Fig.~\ref{fig:studied_systems}c).
Physically, these phases reflect that low-energy quasiparticles (anyons)---present only virtually in the topological phase---either condense (Higgs mechanism~\cite{kogut_lattice_qcd_higgs_mechanism}, transition at fixed $h_x$) or become confined by an effective potential that rises linearly with separation (transition at fixed $h_z$), in both cases destroying topological order.
Importantly, while these trivial phases describe qualitatively very different physics, the Fradkin--Shenker theorem shows that they are adiabatically connected via a path around a first-order transition line extending along the $h_x = h_z$ line (see Fig.~\ref{fig:studied_systems}c)~\cite{fradkin_shenker}.

The extended toric code has recently attracted considerable attention in the context of quantum simulation, as a simple $\mathbb{Z}_2$ lattice gauge theory enabling studies of high-energy phenomena with quantum simulators~\cite{barbiero_lattice_gauge_experiments,halimeh_lattice_gauge_simulators,schweizer_lattice_gauge_experiments}.
It is, however, challenging to identify snapshot-based order parameters for the transition between topological and trivial phases.
Among proposed candidates are the Fredenhagen--Marcu (FM) order parameter~\cite{xu_one_form_symmetries} and percolation-based order parameters~\cite{linsel_percolation,linsel_independent_confinement,dunnweber_percolation_rg}.
The FM estimator involves a ratio of exponentially small quantities, thus demanding large sample sizes for its accurate snapshot-based evaluation; in practice it is reliably computed only from full-state representations (e.g., tensor networks~\cite{xu_one_form_symmetries} or neural quantum states~\cite{kufel_approximately_symmetric_nqs}).
However, even then it may signal unphysical phase transitions~\cite{xu_one_form_symmetries}.
Percolation order parameters are more snapshot-friendly, yet also basis-dependent.
On the self-dual square lattice, in a single basis they readily detect either the deconfined--confined ($x$-basis) or the deconfined--Higgs ($z$-basis) transition~\cite{linsel_percolation}, while the remaining transition is typically much less directly accessible.
It remains to be understood whether both transitions can be robustly detected using single-basis snapshots~\cite{linsel_percolation}.
The non-self-dual triangular lattice hosts \emph{two} distinct types of anyons ($e$- and $m$-anyons), and the topological phase is the one where both are deconfined.
Each basis probes the confinement of one anyon type, so reconstructing the full phase diagram requires combining information obtained independently from basis-specific percolation order parameters~\cite{linsel_independent_confinement}.
At the same time, distance learning naturally combines information from multiple bases into unified $f$-divergences.
It is thus instructive to ask whether it can reproduce the full phase diagram directly from snapshots in both bases.

\subsubsection{Small external fields}
First, we apply distance learning to the phase diagram of the extended toric code at small external fields.
We consider a lattice of size $L = 10$ and sample 195 points in the $(h_x, h_z)$ parameter space, distributed on the $[0, 0.6]\times[0, 0.6]$ square in a $13\times15$ rectangular grid.
For each point we produce 5000 snapshots per each of $x$ and $z$ bases.
When supplied with snapshots from both $x$ and $z$ basis, distance learning correctly finds confinement, Higgs, and first-order transition lines as shown in Fig.~\ref{fig:toric_code_results}a.
However, it does \emph{not} unite the confined and Higgs phases into a single adiabatically connected phase; we study this behaviour further in the next section.
Similarly to the percolation order parameters, when provided with snapshots from one basis, distance learning captures only one of two deconfinement transitions: either to the confined phase ($x$-basis), or to the Higgs phase ($z$-basis) (Fig.~\ref{fig:toric_code_results}b,c).
This underscores the importance of multi-basis measurements for mapping phase diagrams of exotic phases, as well as the ability of distance learning to synthesize such information into a single global quantity.

We also apply distance learning to the extended toric code defined on triangular (2D, $3L^2$ spins) and cubic (3D, $3L^3$ spins) lattices, whose phase diagrams were studied by some of us~\cite{linsel_independent_confinement}.
In both cases we qualitatively reproduce the boundary of the deconfined phase as shown in Fig.~\ref{fig:toric_code_results}e,f.
The clustering also tracks the critical lines as their own thin clusters (crimson and purple in Fig.~\ref{fig:toric_code_results}a,c,e,f), and, for the cubic lattice, we additionally observe a light-blue cluster adjacent to the would-be first-order line (Fig.~\ref{fig:toric_code_results}f).
We view this as a supercritical-like crossover region, in the same spirit as the conjecture discussed below.

\subsubsection{High external fields}
To better understand why distance learning perceives the confined and Higgs phases as two different clusters, we apply it to the region of high external fields (Fig.~\ref{fig:toric_code_results}d).
We expect that the adiabatic connection between the two phases should manifest itself clearly in this region, thus leading to a single cluster being identified by our algorithm.
Surprisingly, distance learning splits the phase diagram into \emph{two} large clusters, separated by wide areas not attributed to any cluster (Fig.~\ref{fig:toric_code_results}d).
We attribute the found clusters to the high-field extensions of the Higgs and confined phases, while the unclustered region is centered around the nominal first-order transition line.
We conjecture that distance learning thus senses the supercritical region around the $h_z = h_x$ line, similar to that found in the liquid--gas phase diagram.
This region is described by the same order parameter as either Higgs or confined phases, yet certain observables peak at the boundaries of this region, thus making it distinguishable at the level of snapshot distributions.
To corroborate this, we plot the sum of $ \sigma^x $ and $\sigma^z$ susceptibilities in Fig.~\ref{fig:toric_code_results}g, and observe that the found unclustered region closely matches the region where these susceptibilities deviate substantially from zero within the observed region.
The location of the found region also coincides with the ``trivial'' region discussed by Xu~\etal{}~\cite{xu_one_form_symmetries} in the context of emergent 1-form symmetries, and it remains to be understood whether these two are the same.
Overall, distance learning not only resolves the topological-to-trivial transitions, but also isolates a supercritical-like crossover region that is not necessarily a distinct thermodynamic phase, yet is clearly distinguished by its structure at the level of snapshot distributions.

\subsection{Fermionic $t$--$J$ model on a triangular lattice}
\subsubsection{Model}
The final application of distance learning is to fermionic systems on a triangular lattice.
Such systems have recently drawn considerable attention thanks to experimental progress in realizing different types of moir\'e systems with van der Waals heterostructures, especially transition metal dichalcogenides (TMDCs).
Recent studies have revealed that quantum magnetism in these materials emerges upon doping the correlated insulator at half-filling~\cite{TMD_Tang2020,TMD_Ciorciaro2023}, signalling the onset of kinetic magnetism. 
This kind of magnetism is not driven by conventional exchange interactions but rather by the motion of doped charge carriers. 
In particular, the emergence of Nagaoka ferromagnetism has been confirmed in these materials~\cite{TMD_Ciorciaro2023}.

These discoveries inspired simulation of the Fermi--Hubbard model on the triangular lattice with ultracold atoms in optical lattices~\cite{Schauss2023,OL_Xu2023,OL_Lebrat2024,OL_Prichard2024}.
Kinetic magnetism and formation of multiparticle bound states have also been explored in experiments with Rydberg arrays~\cite{qiao_browayes_nagaoka}.
Thanks to the single-site resolution available in these experiments, magnetic polarons have been identified through measurements of higher-order correlation functions, revealing that kinetic magnetism is accompanied by the formation of strong higher-order correlations~\cite{OL_Lebrat2024,OL_Prichard2024}.
Analysing such snapshots is thus a natural application for distance learning, designed to capture global features of probability distributions, not limited to correlations of any order.

Specifically, we study snapshots from the fermionic $t$--$J$ model on the triangular lattice, which is an effective theory for Fermi--Hubbard systems in the limit of strong on-site repulsion, and is described by the following Hamiltonian:
\begin{equation}\label{equ:t_j_hamiltonian}
    \begin{gathered}
        \hat{H}_{t-J} = \hat{\mathcal{P}}_G \left[ -t \sum_{\langle i, j \rangle, \sigma} \left( \hat{c}_{i,\sigma}^\dagger \hat{c}_{j,\sigma} + \text{h.c.} \right) \right] \hat{\mathcal{P}}_G+ J \sum_{\langle i, j \rangle} \hat{\mathbf{S}}_i \cdot \hat{\mathbf{S}}_j.
    \end{gathered}
\end{equation}
Here $t$ is the hopping strength, $J$ is the superexchange interaction strength, and $\hat{\mathcal{P}}_G$ is the Gutzwiller projector, which removes all doubly occupied sites. To access the doublon-doped regime, we consider a negative tunneling strength, $t<0$.

\subsubsection{Kinetic ferromagnetism}
\begin{figure*}
	\centering
	\includegraphics[width=\linewidth]{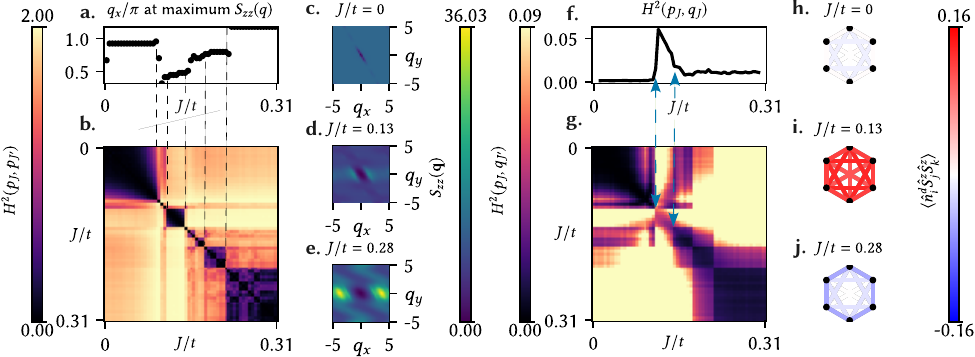}
	\caption{Discriminative distance learning of the doublon-doped triangular-lattice $t$--$J$ model and related diagnostic correlators.
\textbf{a.} Position of the dominant peak in the spin structure factor $S_{zz}(\mathbf{q})$, shown as $q_x/\pi$ at $\arg\max_{\mathbf{q}} S_{zz}(\mathbf{q})$ versus $J/t$.
\textbf{b.} Matrix of pairwise squared Hellinger divergences $H^2(p_J,p_{J^\prime})$ between snapshot distributions at different couplings.
\textbf{c--e.} Representative momentum-space patterns of the spin structure factor $S_{zz}(\mathbf{q})$ for several values of $J/t$ from distinct clusters. The patterns are obtained by evaluating the finite-size Fourier transform on a dense grid in the Brillouin zone to visualise the momentum-space profile and locate its dominant peak.
\textbf{f.} Diagonal divergences $H^2(q_J \Vert p_J)$ between the original plaquette distribution $p_J$ and a surrogate $q_J$.
\textbf{g.} Full matrix of divergences $H^2(q_{J^\prime} \Vert p_J)$ across pairs of different superexchange strengths $(J/t,J^\prime/t)$.
\textbf{h--j.} Connected three-point Nagaoka correlator $\langle \hat{n}^d_i \hat{S}^z_j \hat{S}^z_k\rangle - \langle \hat{n}^d_i \rangle \langle \hat{S}^z_j \hat{S}^z_k \rangle$ evaluated on hexagonal plaquettes extracted from snapshots, shown for representative couplings.}
	\label{fig:nagaoka_results}
\end{figure*}
We first apply distance learning to the $t$--$J$ model doped with doublons on an $18\times3$ cylinder, whose phase diagram was recently discussed in Refs.~\cite{morera_nagaoka,he2023itinerantmagnetismtriangularlattice,Chen2025}.
We consider a slice of the phase diagram at small doping $\delta = 6/54 \approx 0.11$ and superexchange strengths $J/t$ ranging from $0$ to $0.31$.
The matrix of learnt distances shown in Fig.~\ref{fig:nagaoka_results}b suggests the existence of multiple clusters in the phase diagram of this system.
Distance learning offers no further insight into the nature of these clusters, and thus we resort to standard correlation-function-based diagnostics.
We calculate the spin structure factor along the $z$-axis $S_{zz}(\mathbf{q})$ for each $J/t$.
We find that clusters can be readily distinguished by the position of the dominant peak in $S_{zz}(\mathbf{q})$.
Fig.~\ref{fig:nagaoka_results}a and Fig.~\ref{fig:nagaoka_results}c--e corroborate this by showing the peak position as a function of $J/t$ and representative structure factors, respectively.
Thus, distance learning naturally partitions the phase diagram into regions with qualitatively different spin correlations.
It does so without any prior knowledge of the system or its phases and without any inductive bias towards specific correlation functions.

The doped system is also expected to exhibit strong charge-spin-spin correlations, which go beyond the usual second-order spin correlations distinguishing the clusters.
In particular, the system hosts Nagaoka kinetic ferromagnetism at small $J/t$~\cite{morera_nagaoka}---a phase driven by coherent doublon propagation that creates ferromagnetic bubbles around the doublons.
To probe these higher-order correlations, we extract from each snapshot all overlapping hexagonal plaquettes and evaluate the three-point correlator $\langle \hat{n}^d_i \hat{S}^z_j \hat{S}^z_k \rangle$.
Here $\hat{n}^d_i$ is the doublon number operator at site $i$ and $\hat{S}^z_j$ ($\hat{S}^z_k$) is the spin operator at site $j$ ($k$).
We show this correlator at several values of $J/t$ in Fig.~\ref{fig:nagaoka_results}h--j.
It is strongly positive for $J/t\in[0.1,0.15]$, indicating the proliferation of ferromagnetic bubbles, while in the remaining regions these third-order correlations remain weak (an order of magnitude smaller).
This provides a concrete higher-order diagnostic, but it relies on prior knowledge of which correlator to probe.

We therefore introduce the following procedure to assess the role of higher-order correlations using distance learning alone.
For each distribution $p_{J}(\xvec)$ of plaquettes extracted from snapshots at superexchange $J$ we construct a surrogate distribution $q_{J}(\xvec)$ which matches the first- and second-order marginals of $p_{J}(\xvec)$, but is otherwise maximally uncorrelated (see Appendix~\ref{sec:max_entropy_construction} for details).
We then estimate the matrix of squared Hellinger divergences $H^2\!\left(q_{J^\prime}\, \Vert \,p_{J}\right)$ for different pairs of superexchange strengths $J/t$ and $J^\prime/t$.
As shown in Fig.~\ref{fig:nagaoka_results}g, the surrogates are within small Hellinger distances from the original distributions across the phase diagram, except in the window $J/t\in[0.1,0.15]$.
This is most apparent on the diagonal, where $H^2(q_J \Vert p_J)$ peaks sharply (Fig.~\ref{fig:nagaoka_results}f).
We therefore conclude that, within this window, the original and surrogate plaquette distributions differ qualitatively despite having identical first- and second-order marginals, and that this discrepancy is driven by higher-order correlations.
A limitation of this procedure is that it does not pinpoint which higher-order correlators are most relevant. 
Nevertheless, it provides a systematic way to show when low-order correlations do \emph{not} fully dominate the physics.

\subsubsection{Kinetically-induced bound states}
\begin{figure*}
	\centering
	\includegraphics[width=\linewidth]{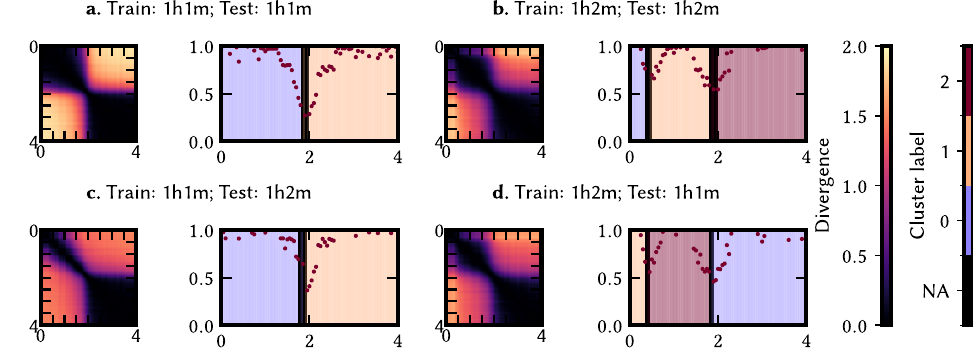}
	\caption{Discriminative distance learning of the hole-doped $t$--$J$ model phase diagram. \textbf{a. (b.)} Matrix of pairwise Hellinger divergences and clustering results for the $t$--$J$ model on a triangular ladder doped with 1 hole and 1 magnon (1 hole and 2 magnons). \textbf{c. (d.)} Surrogate Hellinger divergence matrices and clustering results obtained by applying the discriminator trained on distributions from one doping regime to snapshots from the other doping regime.}
	\label{fig:t_j_results}
\end{figure*}
Finally, we apply distance learning to snapshots from slightly hole-doped $t$--$J$ ladders near full polarisation.
In this regime, kinetically frustrated holes form composite bound states with magnons, leading to strong higher-order correlations~\cite{Batista_attraction,morera_t_j_ladders}. These composite objects have recently been observed experimentally using Rydberg tweezer arrays~\cite{qiao_browayes_nagaoka}.
For convenience, we refer to these composites as distinct ``phases'', even though they are not thermodynamic phases in the conventional sense.

We consider two doping regimes: 1 hole and 1 magnon (1h1m), and 1 hole and 2 magnons (1h2m) (see Fig.~\ref{fig:studied_systems}d).
For $J/t > 1.8$, holes and magnons are completely unbound in both regimes; for $J/t < 1.8$, the hole and a magnon bind into a composite object.
The 1h2m doping regime features an additional bound state in which all three particles bind together for $J/t < 0.5$.
For each doping regime, distance learning yields a block-diagonal $f$-divergence matrix, with the expected two blocks in 1h1m and three blocks in 1h2m (Fig.~\ref{fig:t_j_results}a,b).
Subsequent clustering recovers boundaries that quantitatively match the known binding thresholds.

As a natural next step, it is desirable to compare the origin of phases in different doping regimes.
However, the snapshot distributions for different dopings belong to different particle number sectors, and thus do not overlap.
This leads to trivially maximal $f$-divergences (i.e. $H^2(p_J^{\mathrm{1h1m}} \Vert p_{J^\prime}^{\mathrm{1h2m}}) = 2 \ \forall J, J^\prime $).
As a result, distance learning would perceive each particle number sector as its own cluster family.

We therefore explore a different avenue to relate different dopings.
We note that \emph{any} vector $\xvec$ can be fed into a trained discriminator, including snapshots from a different doping regime.
The discriminator interprets $\xvec$ as a point in a \emph{real} space $\mathbb{R}^L$ (with $L$ being the snapshot length) and maps it to posteriors $\Pr(J/t|\xvec)$ over the couplings it was trained on.
Effectively, it decides which distribution $p_J$ the snapshot is most consistent with.
We therefore perform the following diagnostic: we deliberately feed snapshots $\xvec \sim p^{\mathrm{1h2m}}_{J}$ into a discriminator trained on $\{p^{\mathrm{1h1m}}_{J}\}_{J}$ (and vice versa) and reuse its outputs in the same estimator as before.
This defines a \emph{surrogate} Hellinger divergence matrix across $J/t$.

Remarkably, the surrogate matrices retain a pronounced block structure---two clusters when the 1h1m discriminator is evaluated on 1h2m snapshots, and three clusters in the reverse (Fig.~\ref{fig:t_j_results}c,d).
The resulting HDBSCAN boundaries match the known binding thresholds too.
We suggest the following intuition: one can regard each $p_J$ as a cloud of samples in $\mathbb{R}^L$, while the discriminator is a continuous function on this space.
If, at fixed $J/t$, the 1h2m cloud lies near the corresponding 1h1m cloud, then cross-evaluating the discriminator amounts to a mild perturbation of its usual inputs, retaining the structure of the surrogate Hellinger matrix.
If instead the 1h2m snapshots lie systematically far from the 1h1m training data, the discriminator is probed in a genuinely unseen region. It thus assigns nearly uniform posteriors over $J/t$, washing out structure in the surrogate distance matrix.
We thus view the clear block structure of the surrogate matrices as evidence that $p^{\mathrm{1h1m}}_{J}$ and $p^{\mathrm{1h2m}}_{J}$ distributions remain close in the ``point cloud'' sense not captured by the \emph{bona fide} cross-doping $f$-divergences.

\section{Discussion}\label{sec:discussion}
\subsection{Relation to the previous works}
We find it important to position our approach with respect to existing works on unsupervised information-geometric exploration of phase diagrams from snapshots.

Initial works in this direction focused exclusively on locating phase transitions via self-supervised training of neural networks. 
In ``learning by confusion''~\cite{nieuwenburg_confusion,liu_2d_confusion}, one considers all possible bipartitions of the phase diagram into two assumed phases, trains a classifier for each bipartition, and identifies the transition from the split with the highest classification accuracy. 
Regression-based approaches instead train a network to predict the control parameter $\eta$ from a single snapshot, and then locate the transition from peaks in the susceptibility of the regressed value~\cite{schafer_vector_divergence,greplova_vector_divergence,arnold_vector_divergence} or minima in its variance~\cite{guo_regression_uncertainty}. 
These methods have been shown to detect both conventional symmetry-breaking~\cite{nieuwenburg_confusion,liu_2d_confusion,schafer_vector_divergence} and topological transitions~\cite{greplova_vector_divergence,arnold_vector_divergence}. 
At the time, their indicators were largely heuristic and did not directly yield critical exponents.

Subsequent work made the connection to information geometry explicit.
Ref.~\cite{arnold_connection_to_fisher} related the existing transition indicators from Ref.~\cite{schafer_vector_divergence,greplova_vector_divergence,arnold_vector_divergence} to the square root of the classical Fisher information. 
Ref.~\cite{arnold_f_divergences} then estimated the Fisher information directly by computing $f$-divergences between neighbouring snapshot distributions.
However, it used an autoregressive model to access explicit densities $p_{\etavec}(x)$, rather than estimated their ratios from snapshots via a discriminator. 
Finally, Ref.~\cite{kasatkin_fisher_training} proposed a self-supervised discriminator-based estimator of the classical Fisher information from snapshots alone, at the cost of retraining around every phase diagram point.

In summary, these approaches adopt a largely ``local'' perspective and target transition \emph{location} via Fisher information, typically without pursuing critical scaling.
In contrast, distance learning shifts the focus to a ``global'' perspective: we consider statistical distances as the main object of interest.
This enables us to reconstruct phase diagrams, access the Fisher information, and study its critical scaling within a single unified framework.

\subsection{Future directions and outlook}
We have shown that discriminative distance learning can identify distinct correlation regimes in diverse quantum many-body systems: those with conventional local order parameters, those with topological order, and those characterised by higher-order correlations.
Thanks to its information-geometric footing and its connection to (quantum) Fisher information, this approach also allowed us to extract the critical exponents directly from snapshots---a capability that remains largely unexplored in machine-learning studies of quantum phases.

Distance learning is however not restricted to equilibrium phases of matter.
It naturally applies whenever one has snapshots from quantum states living on a parameterised manifold.
This includes dynamical settings---e.g., quenches---potentially hosting dynamical phase transitions.
There, $f$-divergences relative to an equilibrium state could play the role of the Loschmidt echo, a quantity widely sought after and hard to estimate in quantum simulator experiments~\cite{karch_loschmidt_echo}.
Moreover, $f$-divergences induce generalised entropies and mutual informations beyond the conventional Shannon ones~\cite{polyanskiy_information_theory_textbook}.
It is appealing to explore their snapshot-based estimation and non-equilibrium applications.

While we touched upon interpretability in several case studies---via comparisons to controlled surrogates and via cross-sector generalisation tests, much remains to be explored.
Distance learning only requires correct density-ratio estimates, and is agnostic to the specific discriminator architecture.
A natural next step is thus to replace the generic neural discriminator with a more interpretable discriminator.
This could, for example, take the form of kernel logistic regression with physically motivated kernels~\cite{sadoune_pollet_svm}, or specialised neural networks designed to extract specific correlation functions from snapshots~\cite{bohrdt_correlator_cnn, coles_rydberg_pca,dawid_tetriscnn}.

Several practical questions also arise on the experimental side. 
Empirically, we have observed that robust phase diagrams can already be obtained with a modest sampling budget ($\sim 10^3$ snapshots per point, see Appendix~\ref{sec:nn_architecture_and_training}).
At the same time, estimating Fisher information requires balancing the grid discretisation step.
If it is too large, the finite-difference approximation fails to capture the local geometry.
If it is too small, neighbouring points become close, which increases the statistical noise and reduces learning efficiency.
It is thus important to understand how to sample phase diagram points so that the geometry of the ground-state manifold is faithfully represented.
It is also crucial to have guidance on which measurement bases are required to capture relevant physical regimes, given both limitations of the current experimental setups and the basis-sensitive nature of certain phase transitions.

There is also a set of questions related to the very nature of distance learning.
In its current form, it yields the metric structure of the phase-diagram manifold---it is natural to ask what additional geometric quantities might be accessible. 
In particular, it would be interesting to explore whether multi-basis measurements can be leveraged to define a complex-valued distance (or overlap surrogate) from snapshots, potentially enabling access to Berry phases and Berry curvature.

It is also appealing to connect distance learning to classical shadows, able to estimate overlaps between quantum states using only random single-shot measurements~\cite{elben_randomized_measurements_toolbox}. 
Classical shadows, however, reconstruct density matrix of each state separately, whereas distance learning directly targets the density ratio $p(\xvec)/q(\xvec)$, which is typically an easier statistical task. 
It would be interesting to extend this ``density-ratio'' viewpoint to classical shadows, and to ask whether their rigorous guarantees can be adapted to our setting.

Our study of the doped $t$--$J$ model highlights a further limitation: distance learning is not well suited to comparing phases whose snapshot distributions do not overlap.
While the cross-doping surrogate partially mitigated this issue, it remains an \emph{ad hoc} construction. 
At the same time, the ``point-cloud'' notion of proximity suggested by our cross-evaluation is reminiscent of Wasserstein-type distances, which remain finite even for non-overlapping distributions~\cite{kantorovitch_wasserstein_distance,toth_quantum_wasserstein_distance,neklyudov_wasserstein_monte_carlo}. 
It would be interesting to explore whether this connection can be formalised and exploited in future work, and whether such distances can be estimated reliably from snapshots.

Finally, a promising alternative is to combine our framework with recent machine-learning approaches that learn real-space renormalisation flows and operator content of critical quantum systems~\cite{gokmmen_deep_rg_phase_diagrams,oppenheim_machine_learning_critical_content}. 
This could lead to a fully unsupervised pipeline: discovering the phases and analysing their content based only on information available from projective measurements.

We anticipate that combining information geometry with machine learning adapted to experimental constraints will further prove itself as a valuable tool for exploring and understanding quantum matter.
\section*{Acknowledgments} 
OM thanks Nikita Koritskii for the insightful discussions at the inception stage of this project, as well as Kirill Neklyudov, Andrei Sobolevski, and Alexey Kroshnin for sharing their expertise on information geometry and Wasserstein distances.
The numerical simulations were performed on the Euler cluster operated by the High Performance Computing group at ETH Zürich.
OM, IM, and ED acknowledge support from the SNSF project 200021\_212899, the Swiss State Secretariat for Education, Research and Innovation (contract number UeM019-1), and the ARO grant number W911NF-20-1-0163.
SL, AB, and FG are funded by the Deutsche Forschungsgemeinschaft (DFG, German Research Foundation) under Germany's Excellence Strategy -- EXC-2111 -- 390814868.
This project has received funding from the European Research Council (ERC) under the European Union’s Horizon 2020 research and innovation programme (Grant Agreement no 948141)---ERC Starting Grant SimUcQuam.

\section*{Code availability}
The Python library implementing distance learning, as well as a simple dataset to reproduce the TFIM results is available at~\cite{born2disc_repo}.
We used the open-source library \textsc{ParaToric} to generate the snapshots of the extended toric code~\cite{linsel_paratoric}.
\bibliography{bibliography}

\appendix
\begin{widetext}
\section{Importance sampling derivation}\label{sec:importance_sampling_derivation}
\subsection{Single basis measurements}\label{sec:single_basis_discussion}

Assume that the discriminator was trained on a dataset $\mathcal{X}$ of size $|\mathcal{X}|$, which contains snapshots from a mixture distribution $\pi_p p + \pi_q q$, with $\pi_p$ and $\pi_q$ being the fractions of snapshots coming from distributions $p$ and $q$ respectively ($\pi_p + \pi_q$ = 1).
We use the following importance sampling scheme to estimate the $f$-divergence between distributions $p$ and $q$:
\begin{equation}\label{equ:importance_sampling}
    \fdivarg{q}{p} \approx \frac{1}{|\mathcal{X}|}\sum_{\xvec \in \mathcal{X}} \left[w(\xvec) \cdot f\left(\frac{\pi_p}{\pi_q} \cdot R(\xvec) \right)\right].
\end{equation}
Here $R(\xvec) \coloneqq \frac{\Pr(q|\xvec)}{\Pr(p|\xvec)}$ is the density ratio produced by the discriminator, and $w(\xvec) = \frac{1}{\pi_p (1 + R(\xvec))}$ is the importance weight.
With equal fractions of $p$ and $q$ in the mixture ($\pi_p = \pi_q = \frac{1}{2}$), $R(\xvec) = \frac{q(\xvec)}{p(\xvec)} = r(\xvec)$, as claimed in the main text.
Note that the same configuration \xvec{} might be present in several copies---either being drawn from the same distribution $p(\xvec)$ or after combining snapshots from $p(\xvec)$ and $q(\xvec)$ to form the mixture dataset.
Either way, each copy is considered as a separate sample.

To show the correctness of the proposed scheme, we replace the summation by the mathematical expectation over the mixture distribution.
In other words, we sum over only unique snapshots and replace the factor $\frac{1}{|\mathcal{X}|}$ with the probability of \xvec{} in the mixture distribution $\pi_p p(\xvec) + \pi_q q(\xvec)$:

\begin{equation}\label{equ:importance_sampling_derivation_start}
    \begin{gathered}
    \frac{1}{|\mathcal{X}|}\sum_{\xvec \in \mathcal{X}} \left[w(\xvec) \cdot f\left(\frac{\pi_p}{\pi_q} \cdot R(\xvec) \right)\right] = \sum_{\xvec} \left[ \left(\pi_p p(\xvec) + \pi_q q(\xvec)\right) \cdot w(\xvec) \cdot f\left( \underbrace{\frac{\pi_p}{\pi_q} \cdot R(\xvec)}_{r(\xvec)} \right) \right].
    \end{gathered}
\end{equation}
For the optimal Bayesian classifier the posterior ratio $R(\xvec)$ is equal to $R(\xvec) = \frac{q(\xvec)}{p(\xvec)} \cdot \frac{\pi_q}{\pi_p}$~\cite{devroye_bayes_optimal_textbook}.
Thus, $w(\xvec) = \frac{1}{\pi_p(1 + R(\xvec))} =  \frac{1}{\pi_p + \pi_q \frac{q(\xvec)}{p(\xvec)}} = \frac{p(\xvec)}{\pi_p p(\xvec) + \pi_q q(\xvec)}$ and $\frac{\pi_p}{\pi_q} \cdot R(\xvec) = \frac{\pi_p}{\pi_q} \cdot \frac{q(\xvec)}{p(\xvec)} \cdot \frac{\pi_q}{\pi_p} = \frac{q(\xvec)}{p(\xvec)} \eqqcolon r(\xvec) $.
Substituting these expressions into Eq.~\eqref{equ:importance_sampling_derivation_start} we obtain:
\begin{equation}\label{equ:importance_sampling_derivation_end}
    \begin{gathered}
         \sum_{\xvec} \left[ \left(\pi_p p(\xvec) + \pi_q q(\xvec)\right) \cdot w(\xvec) \cdot f\left(r(\xvec) \right)\right] = \sum_{\xvec} \left[ \left(\pi_p p(\xvec) + \pi_q q(\xvec)\right) \cdot \frac{p(\xvec)}{\pi_p p(\xvec) + \pi_q q(\xvec)} \cdot f\left(\frac{q(\xvec)}{p(\xvec)}\right)\right] = \\ = \sum_{\xvec} p(\xvec) \cdot f\left(\frac{q(\xvec)}{p(\xvec)}\right) \equiv \fdivarg{q}{p}. \qed
    \end{gathered}
\end{equation}

\subsection{Multiple bases measurements}\label{sec:multiple_bases_discussion}
Assume now the dataset contains snapshots coming from different measurement bases.
Suppose the fraction of snapshots coming from the distribution $p$ in the basis $\alpha$ is $\pi_\alpha^{(p)}$, and $\sum_\alpha \pi_\alpha^{(p)} = 1$.
We append the basis label $\alpha$ to each snapshot $\xvec$ and consider joint distributions $p(\xvec, \alpha) = p(\xvec|\alpha) \cdot \pi_\alpha^{(p)}$ and $q(\xvec, \alpha) = q(\xvec|\alpha) \cdot \pi_\alpha^{(q)}$.

We modify the importance sampling estimator~\eqref{equ:importance_sampling} as follows:
\begin{equation}\label{equ:importance_sampling_multiple_bases}
    \fdivarg{q}{p} \approx \frac{1}{|\mathcal{X}|}\sum_{\xvec \in \mathcal{X}}  \left[w(\xvec, \alpha) \cdot f\left(\frac{\pi_p}{\pi_q}  \cdot\frac{\pi_\alpha^{(p)}}{\pi_\alpha^{(q)}}\cdot R(\xvec, \alpha) \right)\right],  
\end{equation}
where $w(\xvec, \alpha) = \frac{1}{\pi_p (1 + R(\xvec, \alpha))}$. 
To show its correctness, we note that the classifier produces the posterior ratio $R(\xvec, \alpha) = \frac{\Pr(q|\xvec, \alpha)}{\Pr(p|\xvec, \alpha)}$ over the joint variable $(\xvec, \alpha)$.
Under the assumption of Bayesian optimality it is equal to:
\begin{equation}\label{equ:posterior_ratio_multiple_bases}
    R(\xvec, \alpha) = \frac{q(\xvec, \alpha)}{p(\xvec,\alpha)} \cdot \frac{\pi_q}{\pi_p} = \frac{q(\xvec|\alpha)}{p(\xvec|\alpha)} \cdot \frac{\pi_\alpha^{(q)}}{\pi_\alpha^{(p)}} \cdot \frac{\pi_q}{\pi_p}.
\end{equation}
Hence, the multiple bases importance weight is equivalent to:
\begin{equation}\label{equ:importance_weight_multiple_bases_derivation}
    w(\xvec, \alpha) = \frac{1}{\pi_p(1 + R(\xvec, \alpha))} = \frac{p(\xvec, \alpha)}{\pi_p p(\xvec, \alpha) + \pi_q q(\xvec, \alpha)}.    
\end{equation}
At the same time the mathematical expectation is now taken over the joint distribution of snapshots and bases $\xvec \sim \pi_p p(\xvec, \alpha) + \pi_q q(\xvec, \alpha)$:
\begin{equation}\label{equ:multuple_bases_importance_sampling_derivation_end}
    \begin{gathered}
    \frac{1}{|\mathcal{X}|}\sum_{\xvec \in \mathcal{X}}   \left[w(\xvec, \alpha) \cdot f\left(\frac{\pi_p}{\pi_q}  \cdot\frac{\pi_\alpha^{(p)}}{\pi_\alpha^{(q)}}\cdot R(\xvec, \alpha) \right)\right] = \\ = \sum_{\alpha} \sum_{\xvec} \left[ \left(\pi_p p(\xvec, \alpha) + \pi_q q(\xvec, \alpha)\right) \cdot  \frac{p(\xvec, \alpha)}{\pi_p p(\xvec, \alpha) + \pi_q q(\xvec, \alpha)} \cdot f\left(\frac{\pi_p}{\pi_q}  \cdot\frac{\pi_\alpha^{(p)}}{\pi_\alpha^{(q)}}\cdot \frac{q(\xvec|\alpha)}{p(\xvec|\alpha)} \cdot \frac{\pi_\alpha^{(q)}}{\pi_\alpha^{(p)}} \cdot \frac{\pi_q}{\pi_p} \right)\right] = \\
    = \sum_\alpha \sum_\xvec p(\xvec, \alpha) f\left(\frac{q(\xvec|\alpha)}{p(\xvec|\alpha)}\right) = \sum_\alpha \pi_\alpha^{(p)} \underbrace{\sum_\xvec \left[ p(\xvec|\alpha) \cdot f\left(\frac{q(\xvec|\alpha)}{p(\xvec|\alpha)}\right) \right]}_{D_f\left(q(\xvec|\alpha) \Vert p(\xvec|\alpha)\right)} = \sum_\alpha \pi_\alpha^{(p)} D_f\left(q(\xvec|\alpha) \Vert p(\xvec|\alpha)\right). \qed
    \end{gathered}
\end{equation}

\end{widetext}

\section{Details of numerical experiments}\label{sec:nn_architecture_and_training}
We follow the standard procedure of training a neural classifier: we split the dataset $\mathcal{X}$ into training proper $\mathcal{X}_{\mathrm{train}}^{\mathrm{TP}}$ and temperature calibration sets $\mathcal{X}_{\mathrm{train}}^{\mathrm{TC}}$ (see below for more details on temperature calibration).
Each of these sets is further divided into training and validation subsets.
We use the posterior probabilities $\Pr(i|\xvec)$ produced by the discriminator, and the true class labels $i$ of the snapshots, to calculate the loss on the training subset $\mathcal{X}_{\mathrm{train}}^{\mathrm{TP}}$:
\begin{equation}\label{equ:cross_entropy_loss}
    \mathcal{L} = -\frac{1}{|\mathcal{X}_{\mathrm{train}}^{\mathrm{TP}}|} \sum_{\xvec \in \mathcal{X}_{\mathrm{train}}^{\mathrm{TP}}} \log \Pr(i|\xvec).    
\end{equation}
We backpropagate the gradients of this loss to update the discriminator parameters, and repeat this procedure for multiple epochs until convergence.
One epoch corresponds to every snapshot \xvec{} being fed into the network once, after which we save the model checkpoint.
We monitor the loss on the validation subset $\mathcal{X}_{\mathrm{val}}^{\mathrm{TP}}$ of samples not seen by the network and stop the training early if the loss did not improve for five epochs.
We then perform temperature calibration on the dataset $\mathcal{X}_{\mathrm{train}}^{\mathrm{TC}}$ as described below.
Finally, we select the model checkpoint with the best value of the validation loss to calculate the $f$-divergence on snapshots from the validation dataset using the estimator~\eqref{equ:importance_sampling}.

\subsection{Discriminator architectures}
Our discriminator consists of two parts.
The first---a ``backbone''---applies a sequence of non-linear operations to the input snapshot $\xvec$ and produces a latent representation \zvec{} of dimension $W$.
The second---a linear classification head---maps \zvec{} to \Npoints{} real numbers known as logits.
Logits are further converted into the posterior probabilities $\Pr(i|\xvec)$ via the softmax function:
\begin{equation}\label{equ:logits_to_posteriors}
    \Pr(i|\xvec) = \frac{e^{l(i|\xvec)}}{\sum_j e^{l(j|\xvec)}}.
\end{equation}

For all systems but 2D Ising model, we use a custom feedforward backbone with $D$ hidden layers.
We use fully-connected layers, which do not impose any inductive biases, and thus should be well-suited to capture non-local correlations of, e.g., systems with topological order.
This is in contrast to, e.g., convolutional layers that assume spatial locality of the relevant correlations.
Each hidden fully-connected layer is followed by the \texttt{LeakyReLU} activation and a custom residual connection: 
\begin{equation}\label{equ:custom_layer}
    \xvec \to \texttt{LeakyReLU}(\mathbf{W} \xvec + \mathbf{b}) + \xvec \eqqcolon \zvec.
\end{equation}
Here $\mathbf{W}$ and $\mathbf{b}$ are learnable weight matrix and bias vector respectively.

In the case of the 2D Ising model, we noticed that with purely feedforward backbones, learnt susceptibility curves do not match the Ferrenberg-Swendsen results well.
Specifically, they decay too quickly away from the phase transition, even though the transition itself is clearly detected.
Therefore, we prepended the feedforward backbone with four convolutional layers, and observed an improved agreement with the Ferrenberg-Swendsen results.
We invite the reader to explore the accompanying repository for the exact details of the considered discriminator architectures~\cite{born2disc_repo}.

\subsection{Robustness to the choice of hyperparameters}
We studied the influence of the training hyperparameters on the learnt distances and phase diagrams.
The hyperparameters include the number of hidden layers $D$ and their sizes $W$, the optimiser learning rate, the training/validation split, etc. 
In Fig.~\ref{fig:1d_tfim_ablations} (see Supplementary Material) we show the fidelity susceptibility curves of the 1D TFIM learnt with different number of snapshots per phase diagram point \Nsnapshots{}.
We observe that even with a very moderate $\Nsnapshots \sim 500$, the learnt curves match theoretical ones.
As $\Nsnapshots$ increases, the match persists. 

In a more extensive study, we correctly reproduce the phase diagram of the extended toric code for a wide range of hyperparameters ($D\sim 3\text{--}7$, $W\sim 64\text{--}256$, $\Nsnapshots  \sim 1000\text{--}5000$) (Fig.~\ref{fig:toric_code_hellinger_ablations}).
We thus find learnt distances to be robust to the choice of the training hyperparameters.

\subsection{Robustness to the choice of $f$-divergence}
We also studied the influence of different choices of $f$-divergences on the learnt distances and phase diagrams (Fig.~\ref{fig:1d_tfim_ablations}--\ref{fig:toric_code_le_cam_ablations}).
Specifically, we consider Le Cam and Jensen-Shannon divergences, which are defined by the following $f$-functions:
\begin{equation}\label{equ:le_cam_definition}
    f_{\mathrm{LeCam}}(t) = \frac{(t - 1)^2}{t + 1},
\end{equation}
\begin{equation}
    f_{\mathrm{JS}}(t) = \frac{1}{2}\left(t \ln t - (t+1) \ln \left(\frac{t + 1}{2}\right)\right).
\end{equation}
We chose these divergences since they yield true metric distances, similarly to the Hellinger distance, and thus are naturally suited for clustering.
We find that inferred phase diagrams and $f$-divergence susceptibilities are robust to the choice of $f$-divergence, in accordance with the common information-geometric nature of these divergences~\cite{polyanskiy_information_theory_textbook}.

\subsection{Temperature calibration of logits}
It is a well-described phenomenon in the deep learning literature that neural discriminators trained with the cross-entropy loss tend to produce overly confident posteriors.
This means that the network predicts $\Pr(p|\xvec) \approx 1$ and $\Pr(q|\xvec) \approx 0$, even when the Bayesian optimal posterior probabilities should be more moderate (i.e. $\Pr(p|\xvec) \approx 0.6$, $\Pr(q|\xvec) \approx 0.4$)~\cite{guo_calibration}.
Since obtaining accurate density ratio estimates is at the core of our method, we apply so-called temperature calibration to the logits once the training is complete~\cite{guo_calibration}.

The temperature calibration procedure introduces a single temperature parameter $T$ and divides the logits by $T$ before applying the softmax function to obtain posteriors:
\begin{equation}\label{equ:temperature_calibration}
    \Pr(i|\xvec) = \frac{e^{l(i|\xvec)/T}}{\sum_j e^{l(j|\xvec)/T}}.
\end{equation}
The initial value of this temperature parameter is $T=1$.
It is subsequently tuned on a small held-out calibration set of samples, by minimising the cross-entropy loss, while keeping the rest of the network parameters fixed.
Once the calibration is over, the posteriors calibrated by $T$ are used to compute the density ratios and $f$-divergences as described in the main text.

\section{Fisher information of thermal states}\label{sec:fisher_information_thermal_states}
To introduce the Fisher information, let us expand an $f$-divergence between two distributions $p_{\eta}$ and $p_{\eta + \Delta \eta}$ to second order in $\Delta \eta$.
To that end, we Taylor expand the function $f(t)$ around $t=1$:
\begin{widetext}
\begin{equation}
\begin{gathered}
    \fdivarg{p_{\eta + \Delta \eta}}{p_{\eta}} = \sum_{\xvec} p_{\eta}(\xvec) f\left(\frac{p_{\eta + \Delta \eta}(\xvec)}{p_{\eta}(\xvec)}\right) = \\ \sum_{\xvec} p_{\eta}(\xvec) \cdot \left[ \underbrace{f(1)}_{0} + f^\prime(1) \cdot \left(\frac{p_{\eta + \Delta \eta}(\xvec)}{p_{\eta}(\xvec)} - 1\right) + \frac{f^{\prime\prime}(1)}{2} \cdot \left(\frac{p_{\eta + \Delta \eta}(\xvec)}{p_{\eta}(\xvec)} - 1\right)^2 + O\left(\Delta \eta^3\right) \right] = \\
    f^\prime(1) \underbrace{\sum_{\xvec} \left(p_{\eta + \Delta \eta}(\xvec) - p_{\eta}(\xvec)\right)}_{1 - 1 = 0} + \frac{f^{\prime\prime}(1)}{2} \sum_{\xvec} p_{\eta}(\xvec) \cdot \left(\underbrace{\frac{p_{\eta + \Delta \eta}(\xvec)}{p_{\eta}(\xvec)} - 1}_{\frac{1}{p(\xvec)}\frac{\partial p_\eta(\xvec)}{\partial \eta} \Delta \eta + O(\Delta \eta^2)}\right)^2 + O\left(\Delta \eta^3\right) = \\
    = \frac{f^{\prime\prime}(1)}{2} \sum_{\xvec} p_{\eta}(\xvec) \left(\frac{1}{p(\xvec)} \cdot \frac{\partial p_\eta(\xvec)}{\partial \eta}\right)^2\Delta \eta^2 + O\left(\Delta \eta^3\right) = \frac{f^{\prime\prime}(1)}{2} \underbrace{\sum_{\xvec} p_{\eta}(\xvec) \left(\frac{\partial \log p_\eta(\xvec)}{\partial \eta}\right)^2}_{I(\eta)} \Delta \eta^2 + O\left(\Delta \eta^3\right).
\end{gathered}
\end{equation}
\end{widetext}
The constant and the first-order terms vanish, and the Fisher information $I(\eta)$ is defined as the prefactor in front of $\frac{f^{\prime\prime}(1)}{2} \Delta \eta^2$, independent of the chosen $f$-divergence.

For Gibbs ensembles, the Fisher information with respect to the inverse temperature $\beta$ has a particularly simple form:
\begin{equation}
        I(\beta) = \mathrm{Var}_{\xvec \sim p_{\beta}}\left[H(\xvec)\right].
\end{equation}
Indeed:
\begin{equation}
    \begin{gathered}
        \partial_{\beta} \log p_{\beta}(\xvec) = \partial_{\beta} \left[ -\beta H(\xvec) - \log Z(\beta) \right] = \\
        = -H(\xvec) - \partial_{\beta} \log Z(\beta).
    \end{gathered}
\end{equation}
Since $\partial_{\beta} \log Z(\beta) = - \langle H \rangle_{\beta}$, we have:
\begin{equation}
    \partial_{\beta} \log p_{\beta}(\xvec) = -H(\xvec) + \langle H \rangle_{\beta}.
\end{equation}
Squaring it and taking the expectation over $p_{\beta}(\xvec)$ we obtain:
\begin{equation}
    I(\beta) = \mathbb{E}_{\xvec \sim p_{\beta}} \left[ \left( -H(\xvec) + \langle H \rangle_{\beta} \right)^2 \right] = \mathrm{Var}_{\xvec \sim p_{\beta}}\left[H(\xvec)\right]. \qed    
\end{equation}
Since we are interested in the Fisher information with respect to temperature $T = \frac{1}{\beta}$, we can use the chain rule to express it as:
\begin{equation}
    I(T) = I(\beta) \cdot \left(\frac{\partial \beta}{\partial T}\right)^2 = \frac{\mathrm{Var}_{\xvec \sim p_{\beta}}\left[H(\xvec)\right]}{T^4}.
\end{equation}
Finally, we use the following relation between the heat capacity and the variance of energy:
\begin{equation}
    C_V = \frac{\partial \langle H \rangle_{\beta}}{\partial T} = \frac{\mathrm{Var}_{\xvec \sim p_{\beta}}\left[H(\xvec)\right]}{T^2}.
\end{equation}
Thus, the Fisher information with respect to temperature can be expressed as:
\begin{equation}
    I(T) = \frac{C_V}{T^2}.
\end{equation}

\section{Na\"ive overlaps between the states}\label{sec:naive_overlaps}
\begin{figure}
	\centering
	\includegraphics[width=0.6\linewidth]{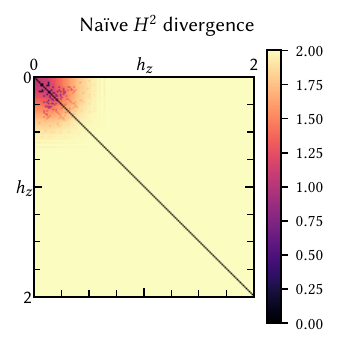}
	\caption{Na\"ively estimated $H^2$ divergence between the ground states of the TFIM model for $L=256$.}
	\label{fig:naive_overlaps}
\end{figure}
Assume one has two batches of snapshots, each containing $\Nsnapshots$ samples, coming from the quantum states $\ket{\psi}$ and $\ket{\phi}$ respectively.
It is possible to na\"ively estimate the phaseless overlap between two quantum states $\ket{\psi}$ and $\ket{\phi}$ as follows:
\begin{equation}
\braket{\phi | \psi} \approx \sum_{\xvec \sim \ket{\psi}} \sum_{\xvec\prime \sim \ket{\phi}} \sqrt{\frac{n_{\ket{\psi}}(\xvec)}{\Nsnapshots}} \sqrt{\frac{n_{\ket{\phi}}(\xvec')}{\Nsnapshots}} \delta_{\xvec, \xvec'}.
\end{equation} 
Here $n_{\ket{\psi}}(\xvec)$ and $n_{\ket{\phi}}(\xvec')$ are the numbers of times the snapshot $\xvec$ and $\xvec'$ appear in the batch produced from $\ket{\psi}$ and $\ket{\phi}$ respectively; the summation is over unique snapshots in each batch; and $\delta_{\xvec, \xvec'}$ is the Kronecker delta.
One can easily convert this overlap estimate into the squared Hellinger divergence estimate via $H^2(\psi, \phi) = 2(1 - \braket{\phi | \psi})$.
We present the results of such na\"ive estimation for the TFIM model in Fig.~\ref{fig:naive_overlaps}.
Apart from a small corner of states deep in the ferromagnetic phase, the na\"ive estimation yields maximal Hellinger divergence for almost all pairs of states, and fails to reproduce the correct phase diagram structure.
This is in line with the expected difficulty of estimating overlaps between high-dimensional probability distributions from a finite number of samples.

\section{Finite size scaling analysis}\label{sec:fss_ansatz}
We fit the standard finite-size scaling ansatz to the experimentally obtained $f$-divergence susceptibility curves to extract the critical exponents of the phase transition:
\begin{equation}
    \chi(\eta, L) / L^d = L^{\alpha_{\mathrm F}/\nu} \cdot f\left( (\eta - \eta_c) \cdot L^{1/\nu} \right).
\end{equation}
To quantify systematic and statistical errors of the extracted critical exponents, we resort to the following procedure.
We consider several different fitting ranges for the phase diagram parameter $\eta$ and system size $L$, and perform the fit for each of these ranges.
For each selected pair of fitting ranges, we bootstrap the data points: for each observed point $\eta$ we sample the value of the susceptibility $\chi$ from a normal distribution with mean and standard deviation equal to the observed value and its error bar respectively.
We perform the fit on the resulting bootstrapped dataset and repeat this procedure 100 times to obtain a distribution of the extracted critical exponents $\alpha_{\mathrm F}$ and $\nu$ for each pair of fitting ranges.
We choose the median of the obtained distribution as the estimate of the critical exponent given the current fitting ranges.
Then we choose the median of such medians across all fitting ranges as the final estimate of the critical exponent.
As the error bars of this estimate, we use 16th and 84th percentiles of the distribution of medians across all fitting ranges.

\section{Ferrenberg-Swendsen histogram reweighting}\label{sec:theoretical_heat_capacity_2d_ising}
\begin{figure}
	\centering
	\includegraphics[width=0.65\linewidth]{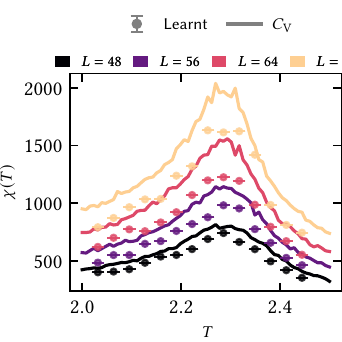}
	\caption{Difference between the finite-difference learnt $f$-divergence susceptibility and the heat capacity in the case of the classical 2D Ising model.}
	\label{fig:2d_ising_heat_capacity_overshoot}
\end{figure}
To benchmark the $f$-divergence susceptibility curves learnt for the 2D classical Ising model, we consider another approach to estimating the density ratios $\frac{p_{T^\prime}(\xvec)}{p_T(\xvec)}$.
Since $p_T(\xvec) = \frac{e^{-\beta H(\xvec)}}{Z_T}$, where $\beta = 1/T$ and $Z_T$ is the partition function at temperature $T$, $p_{T^\prime}(\xvec)/p_T(\xvec)$ can be expressed as:
\begin{equation}
    \frac{p_{T^\prime}(\xvec)}{p_T(\xvec)} = \frac{e^{-\beta^\prime H(\xvec)}}{e^{-\beta H(\xvec)}} \cdot \frac{Z_T}{Z_{T^\prime}} = \frac{Z_T}{Z_{T^\prime}} \cdot e^{-(\beta^\prime - \beta) H(\xvec)}.
\end{equation}
Terms of the form $e^{-\beta H(\xvec)}$ are trivially calculable from a snapshot $\xvec$ and the known 2D Ising Hamiltonian.
At the same time, the ratio of partition functions $\frac{Z_T}{Z_{T^\prime}}$ requires summation over exponentially many $\xvec$.
Nevertheless, in practice it can be estimated to a good accuracy using the Ferrenberg-Swendsen histogram reweighting~\cite{ferrenberg_swendsen_histogram_reweighting_1,ferrenberg_swendsen_histogram_reweighting_2}:
\begin{equation}
    \frac{Z_{T^\prime}}{Z_{T}} = \left\langle e^{\left(-(\beta^\prime - \beta) H \right)} \right\rangle_{T}.
\end{equation}
Indeed:
\begin{equation}
    \begin{gathered}
        \left\langle e^{\left(-(\beta_2 - \beta_1) H \right)} \right\rangle_{\beta_1} = \sum_{\xvec} \frac{e^{-\beta_1 H(\xvec)}}{Z(\beta_1)} e^{\left(-(\beta_2 - \beta_1) H(\xvec) \right)} = \\
        = \frac{1}{Z(\beta_1)} \sum_{\xvec} e^{-\beta_2 H(\xvec)} = \frac{Z(\beta_2)}{Z(\beta_1)}. \qed
    \end{gathered}
\end{equation}
The method provides satisfactory results when $T$ and $T^\prime$ are close, which is the case for our fidelity susceptibility estimation.
For large separation between the distributions it is known to suffer from the exploding variance problem. 
We plug in the density ratios estimated via the histogram reweighting into the estimator~\eqref{equ:importance_sampling}, and then calculate the $f$-divergence susceptibility using the definition in Eq.~\eqref{equ:f_divergence_susceptibility}.

It is also possible to estimate the $f$-divergence susceptibility using its relation to the heat capacity (see Appendix~\ref{sec:fisher_information_thermal_states}).
The latter can be calculated from snapshots as the variance of energy within a batch of samples at a given temperature.
As shown in Fig.~\ref{fig:2d_ising_heat_capacity_overshoot}, the learnt susceptibility curves substantially underestimate the ones obtained from the heat capacity.
We attribute this to a finite $\Delta T$ used to estimate the susceptibility according to Eq.~\eqref{equ:f_divergence_susceptibility}.
Instead, the heat capacities obtained from snapshot ensembles correspond directly to the infinitesimal definition of the susceptibility.
Nevertheless, we suggest that the observed discrepancy is not a fundamental issue, since finite-size scaling performed on the finite-difference estimates collapses the curves onto the expected universal form well.

\section{Maximum entropy construction}\label{sec:max_entropy_construction}
Assume we are given a probability distribution $p(\xvec)$ defined on a discrete configuration space $\xvec = (x_1, x_2, \ldots, x_N)$.
In the context of the $t$--$J$ model plaquettes, N equals 7, and each $x_i$ takes three possible values (without loss of generality we can assume that $x_i=\pm1$ corresponds to the presence of a spin-up or spin-down particle on the site $i$ of the plaquette, while $x_i=2$ corresponds to the presence of a doublon).
We aim to construct a distribution $q(\xvec)$ on the same configuration space, which matches the one- and two-body marginal distributions of $p(\xvec)$ and is otherwise maximally structureless.

This amounts to finding the distribution $q(\xvec)$ that maximises the Shannon entropy $S(q) = -\sum_{\xvec} q(\xvec) \log q(\xvec)$, subject to the following constraints~\cite{mackay_learning_textbook, wainwright_graphical_models}:
\begin{equation}
    \begin{gathered}
        q(x_i\!=\!a) = p(x_i\!=\!a)\quad \forall i \in 1..N,\quad \forall a \in \{-1, +1, 2\} \\
        q(x_i\!=\!a, x_j\!=\!b) = p(x_i\!=\!a, x_j\!=\!b) \quad \forall i, j, \forall a, b \in \{-1, +1, 2\}.
    \end{gathered}
\end{equation}

These constraints can be considered as \emph{expectation value} constraints of the form $\mathbb{E}_{\xvec \sim q} \left[f_k(\xvec)\right] = c_k$.
Here $f_k(\xvec)$ are some functions of the configuration $\xvec$ and $c_k$ are the corresponding expectation values calculated from $p(\xvec)$.
Specifically, the one-body marginal constraints correspond to the functions $f_{i}^{a}(\xvec) = \delta_{x_i, a}$, while the two-body marginal constraints correspond to the functions $f_{i,j}^{a,b}(\xvec) = \delta_{x_i, a} \cdot \delta_{x_j, b}$ ($\delta_{x_i, a}$ is the Kronecker delta).
At the same time, the expectation values $c_k$ are calculated from $p(\xvec)$ as $c_i^a = \sum_{\xvec} p(\xvec) \cdot f_i^a(\xvec)$ and $c_{i,j}^{a,b} = \sum_{\xvec} p(\xvec) \cdot f_{i,j}^{a,b}(\xvec)$. 
The solution to the maximum entropy problem with such constraints is known to be a Gibbs distribution of the form:
\begin{equation}
    q(\xvec) = \frac{1}{Z} e^{-\sum_{i, a} \lambda_i^a f_i^a(\xvec) - \sum_{i, j, a, b} \lambda_{i,j}^{a,b} f_{i,j}^{a,b}(\xvec)}.
\end{equation}
Here $\lambda_i^a$ and $\lambda_{i,j}^{a,b}$ are fitting parameters, similar in spirit to, e.g., chemical potential or temperature in the context of grand canonical ensemble, and $Z$ is the partition function that ensures normalisation of $q(\xvec)$.
In our case, finding the optimal values of these parameters is greatly simplified by the fact that there are in total only $3^7 = 2187$ possible configurations $\xvec$.
Thus, the partition function $Z$ can be calculated via direct summation over all configurations.
We can therefore directly calculate and maximise the entropy $S(q)$ as a function of the fitting parameters $\lambda_i^a$ and $\lambda_{i,j}^{a,b}$.
Such maximisation is known to be a convex optimisation problem, and can be done efficiently using standard gradient-based optimisers~\cite{wainwright_graphical_models}.

\onecolumngrid
\clearpage %

\begin{center}
    \textbf{\large Supplementary material: Distance learning from projective measurements as an information-geometric probe of many-body physics}
\end{center}
\vspace{1em} %
\suppressfloats[t]

\setcounter{section}{0}
\renewcommand{\thesection}{S\arabic{section}}
\renewcommand{\theHsection}{suppsec.\arabic{section}}

\setcounter{figure}{0}
\renewcommand{\thefigure}{S\arabic{figure}}
\renewcommand{\theHfigure}{suppfig.\arabic{figure}}

\setcounter{table}{0}
\renewcommand{\thetable}{S\arabic{table}}
\renewcommand{\theHtable}{supptab.\arabic{table}}

\begin{figure*}[h!]
    \centering

    \subfloat[]{
        \includegraphics[width=0.8\textwidth,height=0.28\textheight,keepaspectratio]{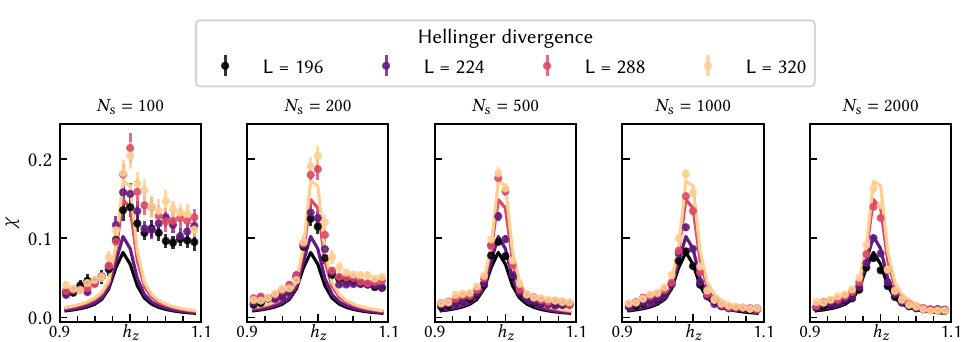}
    }\par\medskip

    \subfloat[]{
        \includegraphics[width=0.8\textwidth,height=0.28\textheight,keepaspectratio]{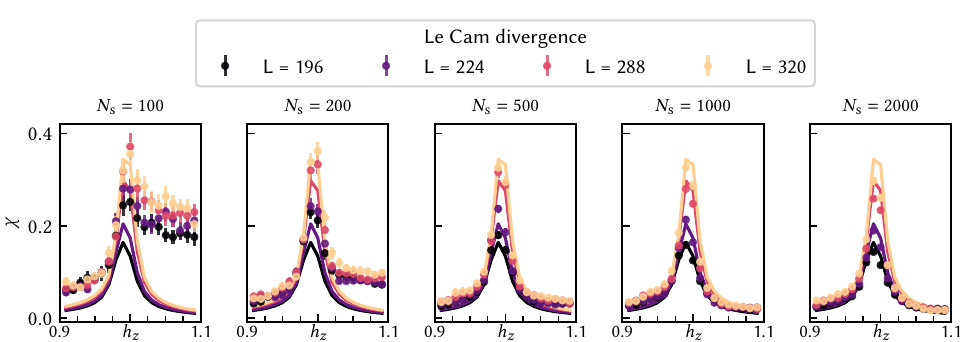}
    }\par\medskip

    \subfloat[]{
        \includegraphics[width=0.8\textwidth,height=0.28\textheight,keepaspectratio]{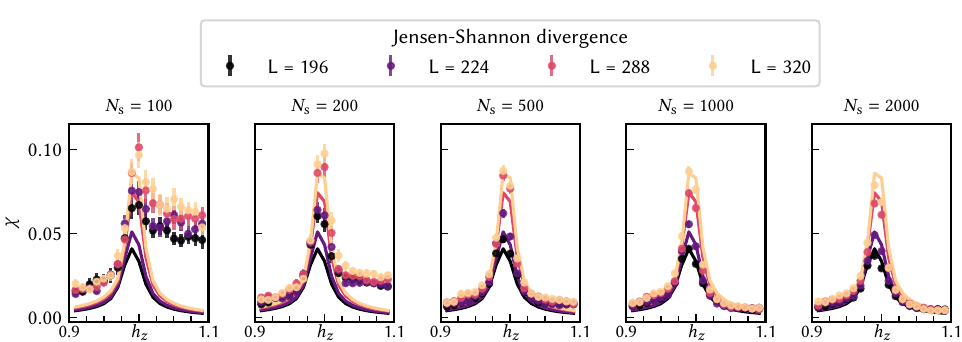}
    }

    \caption{1D TFIM hyperparameter study. We learn the susceptibilities for different numbers of snapshots per phase diagram point $\Nsnapshots$ and different choices of $f$-divergence. We find that the learnt susceptibilities match the DMRG predictions for $\Nsnapshots > 500$ across all $f$-divergence choices.}
    \label{fig:1d_tfim_ablations}
\end{figure*}

\newpage %
\begin{figure*}[t]
    \centering

    \subfloat[]{
        \includegraphics[width=0.4\textwidth,keepaspectratio]{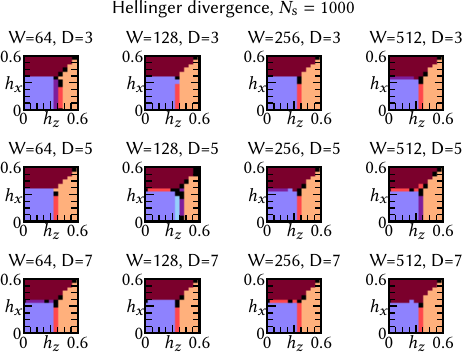}
    }\qquad\qquad
    \subfloat[]{
        \includegraphics[width=0.4\textwidth,keepaspectratio]{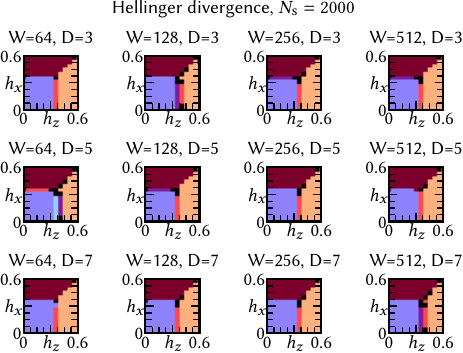}
    }\par\medskip

     \subfloat[]{
        \includegraphics[width=0.4\textwidth,keepaspectratio]{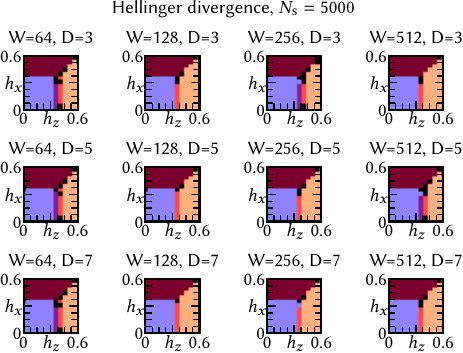}
    }\qquad\qquad
    \subfloat[]{
        \includegraphics[width=0.4\textwidth,keepaspectratio]{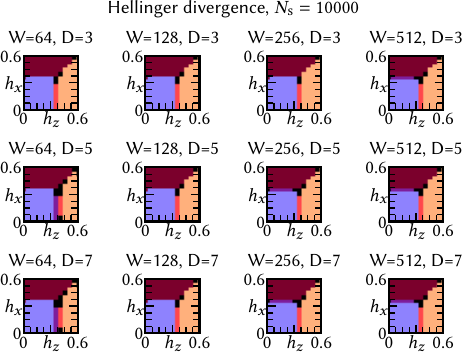}
    }

    \caption{Toric code hyperparameter study for Hellinger divergence. We reconstruct the phase diagram for different numbers of snapshots per phase diagram point $\Nsnapshots$, different network width $W$, and different network depth $D$. We find that the learnt phase diagrams are robust to the choice of hyperparameters: the phase diagram is split into three clusters, with the quantitatively correct phase transition boundaries.}
    \label{fig:toric_code_hellinger_ablations}
\end{figure*}
\begin{figure*}[t]
    \centering

    \subfloat[]{
        \includegraphics[width=0.4\textwidth,keepaspectratio]{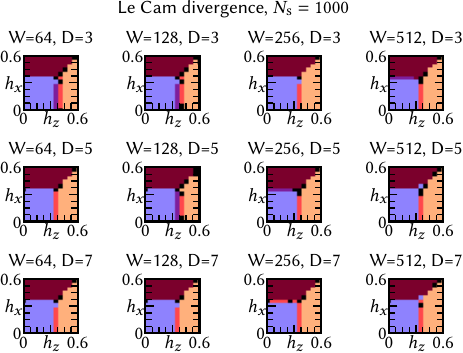}
    }\qquad\qquad
    \subfloat[]{
        \includegraphics[width=0.4\textwidth,keepaspectratio]{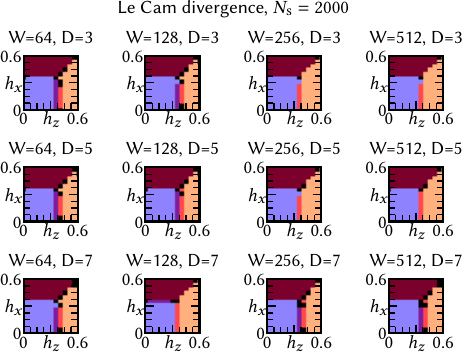}
    }\par\medskip

     \subfloat[]{
        \includegraphics[width=0.4\textwidth,keepaspectratio]{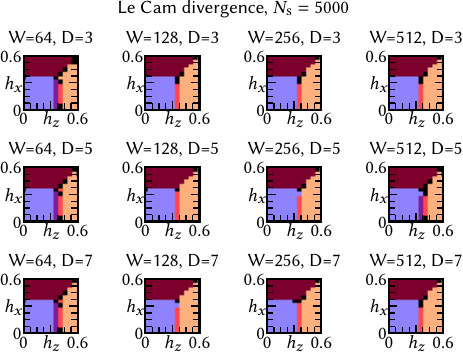}
    }\qquad\qquad
    \subfloat[]{
        \includegraphics[width=0.4\textwidth,keepaspectratio]{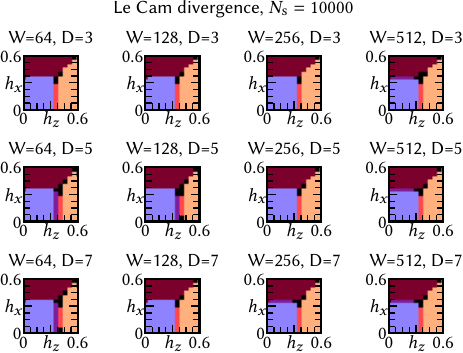}
    }

    \caption{Toric code hyperparameter study for Le Cam divergence. We reconstruct the phase diagram for different numbers of snapshots per phase diagram point $\Nsnapshots$, different network width $W$, and different network depth $D$. We find that the learnt phase diagrams are robust to the choice of hyperparameters: the phase diagram is split into three clusters, with the quantitatively correct phase transition boundaries.}
    \label{fig:toric_code_le_cam_ablations}
\end{figure*}
\begin{figure*}
    \centering

    \subfloat[]{
        \includegraphics[width=0.4\textwidth,keepaspectratio]{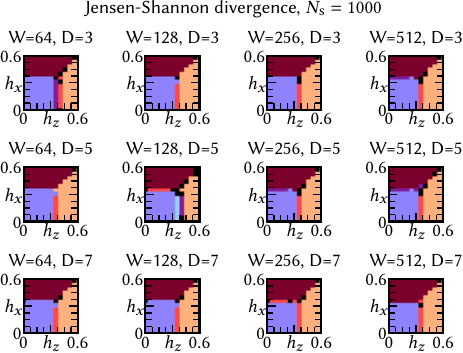}
    }\qquad\qquad
    \subfloat[]{
        \includegraphics[width=0.4\textwidth,keepaspectratio]{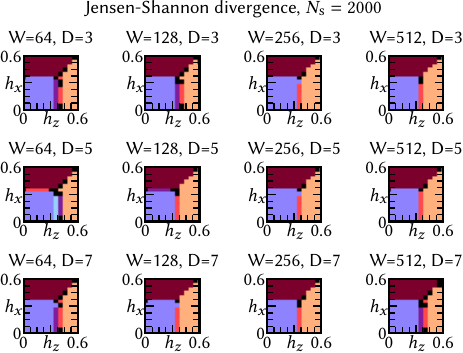}
    }\par\medskip

     \subfloat[]{
        \includegraphics[width=0.4\textwidth,keepaspectratio]{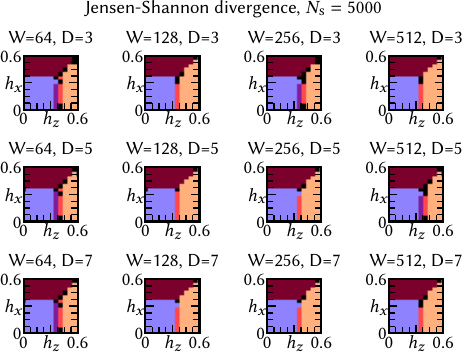}
    }\qquad\qquad
    \subfloat[]{
        \includegraphics[width=0.4\textwidth,keepaspectratio]{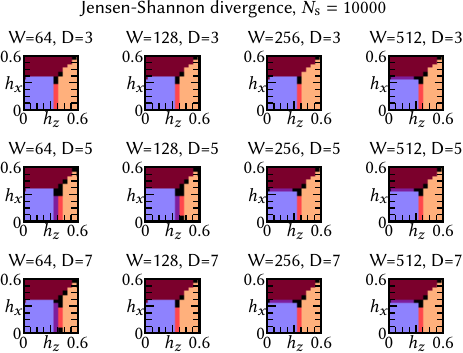}
    }

    \caption{Toric code hyperparameter study for Jensen-Shannon divergence. We reconstruct the phase diagram for different numbers of snapshots per phase diagram point $\Nsnapshots$, different network width $W$, and different network depth $D$. We find that the learnt phase diagrams are robust to the choice of hyperparameters: the phase diagram is split into three clusters, with the quantitatively correct phase transition boundaries.}
    \label{fig:toric_code_jensen_shannon_ablations}
\end{figure*}

\clearpage

\end{document}

%% file: preamble.tex
\ifdefined\preambleImported
\else
    \newcommand{\preambleImported}{}
    \usepackage{graphicx}
    \graphicspath{ {figures/} }
    \usepackage[caption=false]{subfig} %

    \usepackage{amsmath}
    \usepackage{amssymb}
    \usepackage{amsthm}
    \usepackage{mathtools}
    \usepackage{etoolbox}
    \usepackage{dcolumn}  
    \usepackage{multirow}
    \usepackage{mathrsfs}
    \usepackage{braket}
    
    \usepackage[version=4]{mhchem}
    
    \usepackage[noend]{algpseudocode}
    \usepackage{algorithm}

    \usepackage{tabularx}
    
    \newcounter{algosuffix}
    \newcounter{algomain}
    \newcommand{\thealgowithsuffix}{\arabic{algomain}\ifnum\value{algosuffix}>0\Alph{algosuffix}\fi}

    \makeatletter
    \newcommand{\algosuffix}[1]{
      \ifnum\value{algosuffix}=0
        \refstepcounter{algomain} %
      \fi
      \setcounter{algosuffix}{#1}
      \edef\@currentlabel{\thealgowithsuffix}
    }
    
    \makeatother
    
    \usepackage[svgnames, usenames, dvipsnames]{xcolor}
    \usepackage{listings}

    \usepackage{hyperref}
    \hypersetup{
        hidelinks,
        colorlinks=true,
        allcolors=DarkBlue
    }

    \newcommand{\Hhat}{\ensuremath{\hat{H}}}
    
    \newcommand{\etavec}{\ensuremath{\boldsymbol{\eta}}}
    \newcommand{\etaveci}{\ensuremath{\boldsymbol{\eta}^{(i)}}}
    
    \newcommand{\Hhatparam}{\ensuremath{\hat{H}(\etavec)}}

    \newcommand{\Npoints}{\ensuremath{N_{\rm pts}}}    
    \newcommand{\Nsnapshots}{\ensuremath{N_{\rm s}}}

    \newcommand{\fdivarg}[2]{D_f\!\left(#1 \Vert #2\right)}

    \AfterEndEnvironment{proposition}{\noindent\ignorespaces}

    \newcommand{\xvec}{\ensuremath{\mathbf{x}}}
    \newcommand{\zvec}{\ensuremath{\mathbf{z}}}

    \newcommand{\svec}{\ensuremath{\mathbf{s}}}

    \newcommand{\mathexp}{\mathbb{E}}

    \newcommand{\etal}{\textit{et al.}}

\fi